\documentclass{appolb}%
\usepackage{amssymb}
\usepackage{amsmath}
\usepackage{epsfig}
\usepackage{amsfonts}
\usepackage{graphicx}%
\begin{document}

\title{Mesons beyond the quark-antiquark picture}
\author{Francesco Giacosa
\address{Institute of Physics, Jan Kochanowski University, Swietokrzyska 15, 25-406, Kielce, Poland \\
Institute for Theoretical Physics, Johann Wolfgang Goethe University,
Max-von-Laue-Str.\ 1, D--60438 Frankfurt am Main, Germany} }
\maketitle

\begin{abstract}
The vast majority of mesons can be understood as quark-antiquark states. Yet,
various other possibilities exists: glueballs (bound-state of gluons), hybrids
(quark-antiquark plus gluon), and four-quark states (either as
diquark-antidiquark or molecular objects) are expected. In particular, the
existence of glueballs represents one of the first predictions of QCD, which
relies on the nonabelian feature of its structure; this is why the search for
glueballs and their firm discovery would be so important, both for theoretical
and experimental developments. At the same time, many new resonances ($X,Y,$
and $Z$ states) were discovered experimentally, some of which can be well
understood as four-quark objects. In this lecture, we review some basic
aspects of QCD and show in a pedagogical way how to construct an effective
hadronic model of QCD. We then present the results for the decays of the
scalar and the pseudoscalar glueballs within this approach and discuss the
future applications to other glueball states. In conclusion, we briefly
discuss the status of four-quark states, both in the low-energy domain (light
scalar mesons) as well as in the high-energy domain (in the charmonia region)

\end{abstract}

\section{Introduction}

A big part of the Particle Data Group \cite{pdg} contains a list of strongly
interacting and short-living resonances($\tau\sim10^{-22}$ s), whose
properties and decay channels were determined via many experimental
collaborations in the last decades.

Quantum Chromodynamics (QCD) contains only quarks and gluons and is based on
the invariance under local color transformations, denoted as $SU_{c}(3).$
There are 6 types (flavors) of quarks: the quarks $u,$ $d,$ $s$ are light with
$m_{u}=1.8$-$3.0$ MeV$,m_{d}=4.5$-$5.3$ MeV$,$ and $m_{s}=95\pm5$ MeV, while
the quarks $c,b,t$ are heavy with $m_{c}=1.275\pm0.025$ GeV, $m_{b}=$
$4.18\pm0.03$ GeV, and $m_{t}=173.2\pm1.2$ GeV (mass values from \cite{pdg}).
Each quark flavor carries a color: reed, green, or blue. Moreover, there are 8
gluons. Each gluon appears as a color-anticolor state, such as
red-antigreen.\ There is no colorless white gluon, explaining why there are
only 8 of them.

QCD could not be analytically solved in its low-energy domain because
perturbation theory does not apply: this is due to the fact that the running
coupling constant increases for decreasing energy. A related property is
confinement: only `white' states, i.e. states which are invariant under local
$SU_{c}(3)$ transformations, are realized in Nature. Colored objects, such as
quarks and gluons (but also diquarks), are not states which hit our detectors.

Which are then the asymptotic states of QCD? They are called hadrons (hadrons
means `thick' in ancient greek). Two types of hadrons exist: mesons and
baryons. For a proper definition of them, we need to introduce the baryon
number: each quark, independently of its flavor and color, carries a baryon
number $B_{q}=1/3$, while each gluon has a vanishing baryon number, $B_{g}=0$.
Then the following general definitions apply:

\textit{Mesons} are white states (i.e. invariant under color transformations)
which have a vanishing total baryon number. Quark-antiquark states of the type
$\left\vert \overline{q}q\right\rangle $, such as pions, kaons, etc., are
mesons \cite{isgur}. In fact: $B_{\overline{q}q}=B_{q}+B_{\overline{q}}%
=\frac{1}{3}-\frac{1}{3}=0$. They are regarded as \textit{conventional}
mesons, but they are \textit{not} the only possibility \cite{amsler}.
Glueballs are mesonic states made solely of (two or more) gluons and are
denoted as $\left\vert gg\right\rangle .$ They are a long-standing but not yet
fulfilled promise of QCD. Glueballs are mesons because their baryon number is
obviously zero, since each gluon is such. In addition, there are also mesonic
tetraquark states made out of a diquark and an antidiquark, $\left\vert
\left(  \overline{q}\overline{q}\right)  \left(  qq\right)  \right\rangle $,
or mesonic molecular states $\left\vert \left(  \overline{q}q\right)  \left(
\overline{q}q\right)  \right\rangle .$ Hybrids are also outstanding candidates
\cite{hybrid}: they are made of a quark-antiquark couple plus one (or more)
gluon(s), $\left\vert \overline{q}qg\right\rangle .$ In general, each state
made of $n$ quarks, $n$ antiquarks, and $m$ gluons has total baryon number
equal to zero and is in principle a meson. Yet, while quark-antiquark states
were measured in countless experiments, only very recently it was possible to
confirm the existence of mesons beyond the quark-antiquark picture (four-quark
states, either in the tetraquark or molecular-like picture and/or admixtures
of them), see for instance \cite{braaten,besz,beszc0}. For what concerns
glueballs, some candidates exist (especially for the lightest scalar
glueball), but the final verification of their existence has still to come.

\emph{Baryons} are hadronic states with total baryon number $B=1$. Three-quark
states $\left\vert qqq\right\rangle $, such as the neutron and the proton, are
baryons. Also in this case, there are other possible configurations, such as
pentaquarks $\left\vert (qq)(qq)\overline{q}\right\rangle $
(diquark-diquark-antiquark) and molecular-like objects as $\left\vert
(qqq)\left(  \overline{q}q\right)  \right\rangle $, see the recent result in
Ref. \cite{pentaquark}.

In these lectures we concentrate on mesons. In Sec. 2 we review some basic
properties of the QCD Lagrangian (symmetries and large-$N_{c}$) as well as
some general properties of conventional quark-antiquark mesons. In Sec. 3 we
turn our attention to the construction of an effective hadronic model of QCD,
the so-called extended Linear Sigma Model (eLSM). We present its building
blocks in detail since the considerations leading to it are quite general and
are based solely on symmetry. A primary element of the eSLM is the dilaton
field, which is naturally linked to the scalar glueball. Glueballs are then
studied in Sec. 4: the decays and the assignment of the scalar glueball are
presented (the resonance $f_{0}(1710)$ is the most prominent candidate). Then,
the branching ratios of a not-yet discovered pseudoscalar glueball are
calculated. In Sec. 5 we move to four-quark objects. We first study the light
scalar sector, where the resonances $f_{0}(500),$ $f_{0}(980),$ $a_{0}(980),$
and $K_{0}^{\ast}(800)$ are most probably not $\bar{q}q$ states; then we
briefly discuss the status of $X,Y,$ and $Z$ states in the region between
$3$-$5$ GeV, where recent experimental discoveries have nicely shown the
existence of mesons which go beyond the quark-antiquark picture. Finally,
in\ Sec. 6 we present our conclusions.

This work is based on two lectures given at the 55 Cracow School of
Theoretical Physics (Zakopane, 2015). The most important original papers are
Refs. \cite{nf2,susanna,dick,stanilast,psg,psgproc,thomaslast} as well as my
habilitation \cite{hab}.

\section{QCD and its symmetries, mesons, and large-$N_{c}$}

\subsection{Lagrangian of QCD and its symmetries}

As a first step, we write the Lagrangian of $QCD$ for an arbitrary number of
colors $N_{c}$ and quark flavors $N_{f}$ (see, for instance, \cite{weisebook}%
):
\begin{align}
\mathcal{L}_{QCD}  &  =\mathrm{Tr}\left[  \overline{q}_{i}(i\gamma^{\mu}%
D_{\mu}-m_{i})q_{i}-\frac{1}{2}G_{\mu\nu}G^{\mu\nu}\right]  ,\text{ }D_{\mu
}=\partial_{\mu}-ig_{0}A_{\mu}\text{ ,}\nonumber\\
G_{\mu\nu}^{a}  &  =\partial_{\mu}A_{\nu}-\partial_{\nu}A_{\mu}-ig_{0}[A_{\mu
},A_{\nu}],\text{ }A_{\mu}=A_{\mu}^{a}t^{a}\text{ ,} \label{lqcd}%
\end{align}
where $a=1,...,N_{c}^{2}-1$ is the color index, $t^{a}$ are $N_{c}\times
N_{c}$ matrices corresponding to the generators of $SU(N_{c})$ (see below),
$f_{abc}$ are the structure constants of $SU(N_{c})$, and $i=1,...,N_{f}$ is
the flavor index ($N_{f}$ is the number of quark flavors). The part containing
gluons only is called Yang-Mills (YM) Lagrangian:
\begin{equation}
\mathcal{L}_{YM}=\mathrm{Tr}\left[  -\frac{1}{2}G_{\mu\nu}G^{\mu\nu}\right]
=-\frac{1}{4}G_{\mu\nu}^{a}G^{a,\mu\nu}.
\end{equation}
For $N_{c}>1$, the YM Lagrangian contains 3-gluon and 4-gluon vertices. The
gluonic self-interactions are a fundamental property of nonabelian theories,
which is believed to be one of the reasons for the emergence of glueballs.

In Nature, $N_{c}=3$ and $N_{f}=6.$ However, depending on the problem, one can
consider different values for $N_{c}$ and $N_{f}.$ This is why it is useful to
have general expressions.

We now list the symmetries of $\mathcal{L}_{QCD}$ as well as their spontaneous
and explicit breakings:

\bigskip

(i) \textit{Local} color symmetry $SU(N_{c})$.

(ii) Dilatation symmetries and its anomaly (denoted as trace anomaly), i.e.
its breaking trough quantum fluctuations.

(iii) Chiral symmetry $U(N_{f})_{R}\times U(N_{f})_{L}\equiv U(1)_{V}\times
SU(N_{f})_{V}\times U(1)_{A}\times SU(N_{f})_{A}$.

(iv) Axial symmetry $U(1)_{A}$ and the corresponding anomaly (also broken by
quantum fluctuations).

(v) Spontaneous chiral symmetry breaking $SU(N_{f})_{V}\times SU(N_{f}%
)_{A}\rightarrow SU(N_{f})_{V}.$

(vi) Explicit breaking of $U(1)_{A}$ and $SU(N_{f})_{A}$ through nonzero bare
quark masses.

\bigskip

Before we continue, it is important to recall some mathematical properties of
the groups\emph{ }$U(N)$\emph{ }and\emph{ }$SU(N)$, since they appear
everywhere in QCD (both for color and flavor d.o.f.). An element of $U(N)$ is
a complex $N\times N$ matrix fulfilling the following requirement:%
\begin{equation}
U^{\dagger}U=UU^{\dagger}=1_{N}\text{ .}\label{un}%
\end{equation}
One can rewrite $U$ as%
\begin{equation}
U=e^{i\theta_{a}t^{a}}\text{, }a=0,1,...,N^{2}-1\text{ ,}%
\end{equation}
where the matrices $t^{a}$ form a basis of linearly independent $N\times N$
Hermitian matrices. Namely, Eq. (\ref{un}) is in this way automatically
realized. It is usual to set
\begin{equation}
t^{0}=\frac{1}{\sqrt{2N}}1_{N}\text{ .}%
\end{equation}
The other matrices are chosen according to the equation
\begin{equation}
\mathrm{Tr}\left[  t^{a}t^{b}\right]  =\frac{1}{2}\delta^{ab}\text{ with
}a,b=0,1,...,N^{2}-1\text{ , }%
\end{equation}
out of which it follows that%
\begin{equation}
\mathrm{Tr}\left[  t^{a}\right]  =0\text{ for }a=1,...,N^{2}-1\text{ .}%
\end{equation}
A matrix $N\times N$ belongs to the subgroup $SU(N)$ if the following
equations are fulfilled:
\begin{equation}
U^{\dagger}U=UU^{\dagger}=1_{N},\text{ }\det U=1\text{ ;}\label{1csundef}%
\end{equation}
It is clear that a matrix belonging to $SU(N)$ can be written as
$U=e^{i\theta_{a}t^{a}}$ with $a=1,...,N^{2}-1$ (the identity matrix, which is
not traceless, is left out). Then:%
\begin{equation}
\det U=e^{Tr\left[  i\sum_{a=1}^{N^{2}-1}\theta_{a}t^{a}\right]  }=1\text{ .}%
\end{equation}
The matrices $t^{a}$ with $a=1,...,N^{2}-1$ are the generators of $SU(N)$ and
fulfill the algebra:
\begin{equation}
\lbrack t^{a},t^{b}]=if^{abc}t^{c}\text{ with }a,b,c=1,...,N^{2}-1\text{ ,}%
\end{equation}
where $f^{abc}$ are the corresponding antisymmetric structure constants.
Namely, the commutator of two Hermitian matrices is anti-Hermitian and
traceless, therefore it must be expressed as a sum over $t^{a}$ for
$a=1,...,N^{2}-1.$

We remind that for $N=2$ one uses $t^{a}=\frac{\tau^{a}}{2}$ ($a=1,2,3),$
where $\tau^{a}$ are the famous Pauli-matrices, and for $N=3$ one uses
$t^{a}=\frac{\lambda^{a}}{2}$ ($a=1,...,8$), where $\lambda^{a}$ are the
Gell-Mann matrices. These two cases are those which are commonly used in practice.

Finally, we recall also that there is a subgroup of $SU(N),$ denoted as the
center $Z(N)$, whose $N$ elements are given by:
\begin{equation}
Z=Z_{n}=e^{i\frac{2\pi n}{N}}1_{N},\text{ }n=0,1,2,...,N-1\text{ .}
\label{zentrum}%
\end{equation}
Each $Z_{n}$ corresponds to a proper choice of the parameters $\theta_{a}$
(the case $Z_{0}=1_{N}$ corresponds to the simple case $\theta_{a}=0,$ the
other elements to more complicated choices).

\bigskip

We now turn back to the previously listed symmetries of QCD, which we
re-discuss in more detail.

\bigskip

(i) \emph{Local color symmetry }$SU(N_{c}=3)_{c}$

The Yang-Mills fields $A_{\mu}(x)$ is a $N_{c}\times N_{c}$ matrix, $A_{\mu
}(x)=A_{\mu}^{a}(x)t^{a}$, while the quark fields are vectors in color space
($q_{i}^{t}=\left(  q_{i,1},...,q_{i,N_{c}}\right)  ,$ $i=1,...,N_{f}$). Under
$SU(N_{c})_{c}$ local gauge transformations they transform as:
\begin{equation}
A_{\mu}(x)\rightarrow A_{\mu}^{\prime}(x)=U(x)A_{\mu}(x)U^{\dagger}%
(x)-\frac{i}{g_{0}}U(x)\partial_{\mu}U^{\dagger}(x)\text{ , }q_{i}\rightarrow
U(x)q_{i}\text{ },\label{transf}%
\end{equation}
with%
\begin{equation}
U(x)=e^{i\theta_{a}(x)t^{a}}\text{ },\text{ }a=1,...,N_{c}^{2}-1\text{
}(=8\text{ for }N_{c}=3)\text{ .}%
\end{equation}

$\mathcal{L}_{QCD}$ is invariant under (\ref{transf}). Such a symmetry is
automatically fulfilled in each purely hadronic model, since all hadrons
(mesons and baryons) are white, i.e. invariant under a local color
transformation. However, color is also important in hadronic models, since it
is related to the so-called `large-$N_{c}$ limit' in which the group
$SU(N_{c}\gg3)_{c}$ instead of $N_{c}=3$, is considered. Then, although
hadronic fields are invariant under color transformations, the parameters have
special scaling behaviors as function of $N_{c}.$ One can then easily
recognize which parameters are dominant in the large-$N_{c}$ limit, see the
more detailed discussion later on.

At nonzero temperature $T$, the center transformation plays an important role.
Namely, one considers color transformations $U(\tau,\vec{x})$ according to
which $U(0,\vec{x})=1_{N}$ and $U(\tau=\beta=1/T,\vec{x})=Z_{n\neq0}.$ The
$YM$ action at temperature $T$ is invariant under this transformation (often
called center transformation, but care is needed, since it is a peculiar
non-periodic transformation linking two different elements of the center). The
periodicity of the YM fields is still fulfilled ($A_{\mu}\rightarrow A_{\mu
}^{\prime},$ and $A_{\mu}^{\prime}(0,\vec{x})=A_{\mu}^{\prime}(\beta,\vec{x}%
)$). The center symmetry is spontaneously broken in the YM vacuum at high
temperature: the expectation value of the Polyakov loop is the corresponding
order parameter, which is nonzero at high $T$ but is zero at low $T.$ The
nonzero vacuum's expectation value (v.e.v.) of the Polyakov loop signalizes
the deconfinement of gluons.

The center transformation is not a symmetry of the whole $QCD$ at nonzero $T$
(i.e., when quarks are included). Namely, the necessary antisymmetric
condition $q_{i}(0,\vec{x})=-q_{i}(\beta,\vec{x})$ is not fulfilled for the
transformed fields, $q_{i}\rightarrow q_{i}^{\prime}=U(x)q_{i}$, for which
$q_{i}(0,\vec{x})=q_{i}^{\prime}(0,\vec{x})\neq-q_{i}^{\prime}(\beta,\vec
{x})=-Z_{n\neq0}q_{i}(\beta,\vec{x}).$ Thus, center symmetry is explicitly
broken by the quark fields.

\bigskip

(ii) \emph{Dilatation symmetry and trace anomaly }

In the so-called chiral limit $m_{i}\rightarrow0,$ the $QCD$ Lagrangian
contains a single parameter $g_{0},$ which is dimensionless. As a consequence,
QCD is invariant under space-time dilatations,
\begin{equation}
x^{\mu}\rightarrow x^{\prime\mu}=\lambda^{-1}x^{\mu}\, .
\end{equation}
Let us first consider the transformation of the gluon fields
\begin{equation}
A_{\mu}^{a}(x)\rightarrow A_{\mu}^{a\prime}(x^{\prime})=\lambda A_{\mu}%
^{a}(x)\text{ }.
\end{equation}
It is easy to check that the Yang-Mills Lagrangian $\mathcal{L}_{YM}=-\frac
{1}{4}G_{\mu\nu}^{a}G^{a,\mu\nu}$ transforms as $\mathcal{L}_{YM}%
\rightarrow\lambda^{4}\mathcal{L}_{YM},$ then the classical action is
invariant. The corresponding conserved (Noether) current is:
\begin{equation}
J^{\mu}=x_{\nu}T^{\mu\nu}\rightarrow\partial_{\mu}J^{\mu}=T_{\mu}^{\mu
}=0\text{ ,}%
\end{equation}
where the energy-momentum $T^{\mu\nu}$ tensor of the YM-Lagrangian is given
by
\begin{equation}
T^{\mu\nu}=\frac{\partial\mathcal{L}_{YM}}{\partial(\partial_{\mu}A_{\rho}%
)}\partial^{\nu}A_{\rho}-g^{\mu\nu}\mathcal{L}_{YM}\text{ }%
+\text{`symmetrization'.}%
\end{equation}
Quantum fluctuations of gluons break dilatation symmetry \cite{migdal}. As a
consequence, the dimensionless coupling constant $g_{0}$ becomes an
energy-dependent running coupling $g(\mu),$ where $\mu$ is the energy scale at
which the coupling is probed (for instance, in scattering processes, it is
proportional to the energy in the center of mass):
\begin{equation}
g_{0}\overset{\text{renormalization}}{\rightarrow}g(\mu)\text{ .}%
\end{equation}
Then, the following equation arises:
\begin{equation}
\partial_{\mu}J^{\mu}=T_{\mu}^{\mu}=\frac{\beta(g)}{4g}G_{\mu\nu}^{a}%
G^{a,\mu\nu}\neq0\text{ , }\beta(g)=\mu\frac{\partial g}{\partial\mu}\text{ ,}
\label{sa}%
\end{equation}
where it is visible that $\partial_{\mu}J^{\mu}\neq0$ as soon as $\partial
g/\partial\mu\neq0$: dilatation symmetry is \textit{explicitly} broken. The
quantity $\beta(g)$ is the so-called $\beta$-function of the YM theory.
Indeed, if $g$ were constant ($g=g_{0}$), then $\partial_{\mu}J^{\mu}=0,$ but
this is not the case. Namely, already the one-loop level, one has:%
\begin{equation}
\beta(g)=\mu\frac{\partial g}{\partial\mu}=-bg^{3}<0\text{ , }b=\frac{11N_{c}%
}{48\pi^{2}}\text{ .} \label{41}%
\end{equation}
The solution of Eq. (\ref{41}) is:
\begin{equation}
g^{2}(\mu)=\frac{g_{\ast}^{2}}{1+2bg_{\ast}^{2}\log\frac{\mu}{\mu_{\ast}}%
}\text{ }. \label{gmu}%
\end{equation}
The fact that $\beta(g)<0$ explains asymptotic freedom: the coupling $g(\mu)$
becomes smaller for increasing $\mu$ (Nobel 2004). On the other side, for
small $\mu$, the coupling $g(\mu)$ increases. A (not yet analytically proven)
consequence is `confinement': gluons (and quarks) are confined in white
hadronic states.

Eq. (\ref{gmu}) has a (so-called Landau) pole
\begin{equation}
\mu_{pole}=\Lambda_{Landau}=\Lambda_{YM}=\mu_{\ast}e^{\frac{-1}{2bg_{\ast}%
^{2}}}\text{ ,}%
\end{equation}
then%
\begin{equation}
g^{2}(\mu)=\frac{1}{2b\log\frac{\mu}{\Lambda_{YM}}}\text{ .}
\label{gmurunning}%
\end{equation}
Notice also that Eq. (\ref{gmu}) scales as follows in the large-$N_{c}$
limit:
\begin{equation}
g\propto1/\sqrt{N_{c}}\text{ .}%
\end{equation}
This is the starting point of the study of the large-$N_{c}$ limit that we
will describe later on.

Obviously, perturbation theory breaks down when the coupling becomes large. I
t does not mean that $g(\mu=\Lambda_{L})$ becomes infinite, but it means that
at the energy scale $\mu\sim\Lambda_{YM}$ the YM theory (and so whole QCD)
becomes strongly coupled. $\Lambda_{YM}$ cannot be obtained theoretically
because the value $g_{\ast}$ at a certain given $\mu_{\ast}$ (as for instance
the grand unification energy scale $\mu_{\ast}=10^{16}$ GeV) is a priori
unknown. However, the discussion shows a central point: a dimension has
emerged! This is the so-called `dimensional transmutation'.\ Numerically, it
turns out that $\Lambda_{QCD}\simeq250$ MeV: this number affects all hadronic processes.

A \textit{purely nonperturbative} consequence of the scale anomaly is the
emergence of a gluon condensate. Namely, the vacuum's expectation value of the
trace anomaly does not vanish (for $N_{c}=3$):%
\begin{equation}
\left\langle T_{\mu}^{\mu}\right\rangle =-\left\langle \frac{11N_{c}}{48}%
\frac{\alpha_{s}}{\pi}G_{\mu\nu}^{a}G^{a,\mu\nu}\right\rangle \sim
-\frac{11N_{c}}{48}\left(  350\text{-}600\text{ MeV}\right)  ^{4}\text{,}%
\end{equation}
where $\alpha_{s}=g^{2}/(4\pi).$ (The numerical results were obtained via
lattice and sum rules calculations, see Ref. \cite{gluoncondensat} and refs.
therein). Indeed, the quantity $\frac{1}{2}G_{\mu\nu}^{a}G^{a,\mu\nu}$ can be
also expressed as (in the Coulomb gauge)%
\begin{equation}
\frac{1}{2}G_{\mu\nu}^{a}G^{a,\mu\nu}=\frac{1}{2}\left(  \vec{E}^{a}\cdot
\vec{E}^{a}-\vec{B}^{a}\cdot\vec{B}^{a}\right)  \text{ ,}%
\end{equation}
where $\vec{E}^{a}$ and $\vec{B}^{a}$ are the electric and magnetic color
fields, respectively \cite{diakonov}. Perturbation theory shows that at each
order $\left\langle \vec{E}^{a}\cdot\vec{E}^{a}\right\rangle =\left\langle
\vec{B}^{a}\cdot\vec{B}^{a}\right\rangle ,$ thus $\left\langle G_{\mu\nu}%
^{a}G^{a,\mu\nu}\right\rangle $ should vanish accordingly. However, the
existence of nonperturbative solutions such as instantons shows that this
perturbative prediction does \textit{not} hold and a nonzero gluon condensate
is one of the main features of Yang-Mills theory.

As a last step, we add quarks and consider whole QCD. By taking into account
that they have dimension $3/2,$ they transform as
\begin{equation}
q^{\prime}(x^{\prime})=\lambda^{3/2}q(x)\text{ .}%
\end{equation}
In the chiral limit, the discussion is similar upon modifying Eq. (\ref{41})
as:
\[
b=\frac{11N_{c}-2N_{f}}{48\pi^{2}}%
\]
Notice that $b>0$ if $N_{f}<\frac{11}{2}N_{c}$. This condition is fulfilled in
Nature, even in the extreme limit in which all six flavors are taken into
account $(N_{f}=6$ and $N_{c}=3).$ A further explicit breaking of dilatation
symmetry emerges from the nonzero bare quark masses, see the point (vi) below.

In conclusion, the breaking of scale invariance is a very important and deep
phenomenon of QCD. An effective description of QCD should contain this
feature. Moreover, the related concept of a condensate of gluons naturally emerges.

(iii) \emph{Chiral Symmetry }$U(N_{f})_{R}\times U(N_{f})_{L}$

In the chiral limit $m_{i}\rightarrow0$ the Lagrangian $\mathcal{L}_{QCD}$ is
invariant under transformations of the group $U(N_{f})_{R}\times U(N_{f})_{L}%
$. First, we recall that this transformations amount to transforming the
right-handed and left-handed parts of the quark fields separately:%
\begin{equation}
q_{i}=q_{i,R}+q_{i,L}\rightarrow U_{R,ij}q_{j,R}+U_{L,ij}q_{j,L}\text{ ,}
\label{chiralt}%
\end{equation}
with $U_{R}\in U(N_{f})_{R}$ $,$ $U_{L}\in U(N_{f})_{L}$ . We also remind that
the right-handed spinor $q_{i,R}$and left-handed spinor $q_{i,L}$ are defined
as \cite{weisebook}:%
\begin{align}
q_{i,R}  &  =P_{R}q_{i}\text{ , }q_{i,R}^{\dag}=q_{i}^{\dagger}P_{R},\text{
}\overline{q}_{i,R}=\overline{q}_{i}P_{L}\text{ ,}\\
\text{ }q_{i,L}  &  =P_{L}q_{i}\text{ , }q_{i,L}^{\dag}=q_{i}^{\dagger}%
P_{L},\text{ }\overline{q}_{i,L}=\overline{q}_{i}P_{R}\text{ ,}%
\end{align}
with $P_{R}=\frac{1}{2}\left(  1+\gamma_{5}\right)  $ , $P_{L}=\frac{1}%
{2}\left(  1-\gamma_{5}\right)  $. This is the famous chiral symmetry of QCD.

\bigskip

(iv) \emph{Axial transformation} $U(1)_{A}$\emph{ and its anomaly}

The axial transformation $U(1)_{A}$ is a subgroup of $U(N_{f})_{L}\times
U(N_{f})_{R}$ corresponding to the choice
\begin{equation}
U_{A}^{(1)}=U_{R}=U_{L}^{\dagger}=e^{i\nu t^{0}}\text{ ,}%
\end{equation}
that is:
\begin{equation}
q_{i,R}\rightarrow e^{\frac{i\nu}{\sqrt{2N_{f}}}}q_{i,R}\text{ , }%
q_{i,L}\rightarrow e^{\frac{-i\nu}{\sqrt{2N_{f}}}}q_{i,L}\Rightarrow
q\rightarrow e^{i\nu t^{0}\gamma_{5}}q\text{ .}%
\end{equation}
This symmetry is also broken by quantum fluctuations (axial anomaly). The
divergence of the corresponding Noether current
\begin{equation}
A_{\mu}^{0}=\overline{q}\gamma^{\mu}\gamma^{5}q=\sum_{i=1}^{N_{f}}\overline
{q}_{i}\gamma^{\mu}\gamma^{5}q_{i}%
\end{equation}
is nonzero:%
\begin{equation}
\partial^{\mu}A_{\mu}^{0}=-\frac{g^{2}N_{f}}{32\pi^{2}}G_{\mu\nu}^{a}\tilde
{G}^{a,\mu\nu}\neq0\text{ , }\tilde{G}^{a,\mu\nu}=\frac{1}{2}\varepsilon
^{\mu\nu\rho\sigma}G_{\rho\sigma}^{a}\text{ .}%
\end{equation}
Effective models should also display this feature, since it is important for
the description of the pseudoscalar mesons $\eta$ and $\eta^{\prime}.$

(v) \emph{Spontaneous symmetry breaking of chiral symmetry: }$SU(N_{f}%
)_{R}\times SU(N_{f})_{L}\rightarrow SU(N_{f})_{V}$

Spontaneous breaking of chiral symmetry is on of the central properties of the
hadronic world. It explains why pions are so light and why their interaction
is so small when they are slow. It also explains the mass differences between
multiplets and affects their decays.

First, we rewrite the group $U(N_{f})_{R}\times U(N_{f})_{L}$ as follows:
\begin{equation}
U(N_{f})_{R}\times U(N_{f})_{L}\equiv U(1)_{V}\times SU(N_{f})_{V}\times
U(1)_{A}\times SU(N_{f})_{A}\text{ .}%
\end{equation}
$U_{V}(1)$ corresponds to
\begin{equation}
U_{1}=U_{L}=U_{R}=e^{i\theta t^{0}}\text{,}%
\end{equation}
$SU(N_{f})_{V}$ to
\begin{equation}
U_{V}=U_{L}=U_{R}=e^{i\theta_{a}^{V}t^{a}}\text{ }(a=1,...,N_{f}^{2}-1)\text{
,}%
\end{equation}
and $SU(N_{f})_{A}$ to
\begin{equation}
U_{A}=U_{L}=U_{R}^{\dagger}=e^{i\theta_{a}^{A}\lambda^{a}}\text{
}(a=1,...,N_{f}^{2}-1)\text{ .}%
\end{equation}
Note, $SU(N_{f})_{A}$ is not a group since the product of two elements of the
set is not an element of the set. $SU(N_{f})_{A}$ is the transformation set
which is spontaneously broken.

In the chiral limit, the conserved Noether currents corresponding to
$U(N_{f})_{V}\equiv U(1)_{V}\times SU(N_{f})_{V}$ are given by ($a=0,..N_{f}%
^{2}$)
\begin{equation}
V_{\mu}^{a}=\overline{q}\gamma^{\mu}t^{a}q\text{ }\rightarrow\partial^{\mu
}V_{\mu}^{a}=0
\end{equation}
while those corresponding to $SU(N_{f})_{A}$ by ($a=1,..N_{f}^{2}$):%
\begin{equation}
A_{\mu}^{a}=\overline{q}\gamma^{\mu}\gamma^{5}t^{a}q\text{ }\rightarrow
\partial^{\mu}A_{\mu}^{a}=0\text{ .}%
\end{equation}

It turns out that the QCD-vacuum $\left\vert 0_{QCD}\right\rangle $ is not
invariant under the $SU(N_{f})_{A}$ transformation.\ In particular, it means
that the axial charges%
\begin{equation}
Q^{a}=%
{\displaystyle\int}
d^{3}xA_{\mu=0}^{a}%
\end{equation}
do not annihilate the vacuum: $Q^{a}\left\vert 0_{QCD}\right\rangle \neq0.$
Then, according to the Goldstone theorem, the pions emerge as (quasi-)massless
Goldstone bosons. Namely, by considering that $\left[  H_{QCD},Q^{a}\right]
=0$ and by applying this commutator on the vacuum, we get:%
\begin{equation}
0=\left[  H_{QCD},Q^{a}\right]  \left\vert 0_{QCD}\right\rangle =H_{QCD}%
\left(  Q^{a}\left\vert 0_{QCD}\right\rangle \right)  =0\left(  Q^{a}%
\left\vert 0_{QCD}\right\rangle \right)  \text{ .}%
\end{equation}
Then, $Q^{a}\left\vert 0_{QCD}\right\rangle $ is proportional to a massless
state. These are the Goldstone bosons: pions, kaons, and the $\eta$-meson for
$N_{f}=3.$

(vi)\emph{ Explicit symmetry breaking due to nonzero quark masses}

The QCD mass term
\begin{equation}
\mathcal{L}_{mass}=\sum_{i=1}^{N_{f}}m_{i}\overline{q}_{i}q_{i}%
\end{equation}
breaks explicitly many of the aforementioned symmetries.

Being quark masses dimensional, the divergence of the dilatation current
acquires an additional term:
\begin{equation}
T_{\mu}^{\mu}=\frac{\beta(g)}{4g}G_{\mu\nu}^{a}G^{a,\mu\nu}+\sum_{i=1}^{N_{f}%
}m_{i}\overline{q}_{i}q_{i}\text{ .}%
\end{equation}

The symmetry under $U(N_{f})_{V}\equiv U(1)_{V}\times SU(N_{f})_{V}$ is still
valid only if $m_{1}=m_{2}=....=m_{N_{f}},$ but, as soon as mass differences
are present, the divergences of the vector currents $V_{\mu}^{a}=\overline
{q}\gamma^{\mu}t^{a}q$ are in general nonvanishing:
\begin{equation}
\partial^{\mu}V_{\mu}^{a}=i\bar{q}\left[  \hat{m},t^{a}\right]  q\neq0\text{
,}%
\end{equation}
with $\hat{m}=diag\{m_{1},m_{2},...,m_{N_{f}}\}.$ The symmetry $U_{V}(1)$
corresponds to $a=0$, therefore it is always valid independently on the masses
(as it must, being the conservation of the baryon number). In low-energy QCD
it is common to set $m_{1}=m_{u}=m_{d}=m_{2},$ therefore isospin symmetry
$U(N_{f}=2)_{V}$ still holds.

The symmetry under $U_{A}(N_{f})\equiv U_{A}(1)\times SU_{A}(N_{f})$ is broken
as soon as $m_{i}\neq0.$ In fact, the currents $A_{\mu}^{a}=\overline{q}%
\gamma^{\mu}\gamma^{5}t^{a}q$ acquire the divergences%
\begin{equation}
\partial^{\mu}A_{\mu}^{a}=i\bar{q}\gamma^{5}\{\hat{m},t^{a}\}q\neq0\text{ for
}a\neq0\text{ .}%
\end{equation}
The small but nonzero quark masses are responsible of the fact that the pions
are not exactly Goldstone bosons and therefore are not exactly massless.

For the case $a=0$ there are two terms, one from the axial anomaly and one
arising from the nonzero values of masses:%
\begin{equation}
\partial^{\mu}A_{\mu}^{0}=2\bar{q}i\hat{m}\gamma^{5}q-\frac{g^{2}N_{f}}%
{32\pi^{2}}G_{\mu\nu}^{a}\tilde{G}^{a,\mu\nu}\text{ .}%
\end{equation}

For $N_{f}=2$, the explicit breaking through quark masses is small,
$m_{u,d}\ll\Lambda_{YM}.$ For $N_{f}=3$ the mass of $s$ is about $100$ MeV and
is thus of the same order of $\Lambda_{YM}.$ In turn, the explicit breaking
induced by the $s$ quark is in general non-negligible. The other quark flavors
($c,b,t$) are heavy: the breaking of symmetry due to their masses is dominant.
This is why one considers the light-quark sector and the heavy-quark sector separately.

As a last point we mention a result which is usually obtained in the framework
of the Nambu-Jona-Lasinio (NJL) model \cite{njl}, which connects (v) and (vi).
The spontaneous breaking of $SU(N_{f})_{A}$ is also responsible for the
generation of an effective (or constituent) quark mass. For $N_{f}=2$:
\begin{equation}
m\simeq5\text{ MeV}\rightarrow m^{\ast}\simeq300\text{ MeV}\gg m\text{ .}%
\end{equation}
It is now evident that $SU(N_{f})_{A}$ is not a symmetry any longer, since
$\partial^{\mu}A_{\mu}^{a}\propto\{m^{\ast},t^{a}\}\neq0.$ Effective quarks
are quasi-particles which emerge when bare quarks are dressed by gluon clouds.
Notice that analogous results hold also when more advanced approaches are used
to study the quark propagator, see for instance the Dyson-Schwinger study of
Ref. \cite{fischerquark}.

A similar phenomenon takes place also for gluons, although the discussion is
much more subtle because of gauge invariance. Nevertheless, gluons dressed by
gluonic fluctuations also develop an effective mass of about $500$-$1000$ MeV
\cite{binosi,fischergluon}.

\bigskip

\subsection{Mesons}

Quarks and gluons are not the physical states that we measure. They are
confined into hadrons, i.e. mesons (integer spin) and baryons (semi-integer spin).

A \textit{conventional meson} is a meson constructed out of a quark and an
antiquark. Although it represents only one of (actually infinitely many)
possibilities to build a meson, the vast majority of mesons of the PDG can be
correctly interpreted as belonging to a quark-antiquark multiple \cite{pdg}
(see also the results of the quark model \cite{isgur}).

Mesons can be classified by their spatial angular momentum $L$, their spin
$S,$ their total angular momentum $J$ (with $\vec{J}=\vec{L}+\vec{S})$, by
parity $P,$ and by charge conjugation $C$ (summarized by $J^{PC}$). We remind
that $P$ and $C$ are calculated as:%
\begin{equation}
P=(-1)^{L+1}\text{ , }C=(-1)^{L+S}\text{ .}%
\end{equation}

The lightest mesons are pseudoscalar states with $L=S=0\rightarrow
J^{PC}=0^{-+}.$ As explained above, the pions and the kaons are pseudoscalar
(quasi-)Goldstone bosons emerging upon the spontaneous symmetry breaking of
chiral symmetry. As an example, we write down the wave function for the pionic
state $\pi^{+}$ and for the kaonic state $K^{+}$ (radial, spin, flavor,
color):
\begin{align}
\left\vert \pi^{+}\right\rangle  &  =\left\vert n=1\right\rangle \left\vert
L=0\right\rangle \left\vert S=0(\uparrow\downarrow-\downarrow\uparrow
)\right\rangle \left\vert u\bar{d}\right\rangle \left\vert \bar{R}R+\bar
{G}G+\bar{B}B\text{ }\right\rangle \text{ ,}\\
\left\vert K^{+}\right\rangle  &  =\left\vert n=1\right\rangle \left\vert
L=0\right\rangle \left\vert S=0(\uparrow\downarrow-\downarrow\uparrow
)\right\rangle \left\vert u\bar{s}\right\rangle \left\vert \bar{R}R+\bar
{G}G+\bar{B}B\text{ }\right\rangle \text{ .}%
\end{align}
For $L=0,$ $S=1$ one constructs the vector mesons, such as the $\rho$-meson:%
\begin{equation}
\left\vert \rho^{+}\right\rangle =\left\vert n=1\right\rangle \left\vert
L=0\right\rangle \left\vert S=1(\uparrow\downarrow+\downarrow\uparrow
)\right\rangle \left\vert u\bar{d}\right\rangle \left\vert \bar{R}R+\bar
{G}G+\bar{B}B\text{ }\right\rangle \text{ .} \label{rho}%
\end{equation}
For $L=S=1$ one has three multiplets: tensor mesons with $J^{PC}=2^{++},$
axial-vector mesons with $J^{PC}=1^{++},$ and scalar mesons with
$J^{PC}=0^{++}$. By further increasing $L,$ and/or the radial quantum number
$n$, and by including other quark flavours (such as the charm quark) one can
obtain many more multiplets of conventional quark-antiquark states, see Ref.
\cite{pdg}. For instance, the renowned $j/\psi$ meson reads:%
\begin{equation}
\left\vert j/\psi\right\rangle =\left\vert n=1\right\rangle \left\vert
L=0\right\rangle \left\vert S=1(\uparrow\downarrow+\downarrow\uparrow
)\right\rangle \left\vert c\bar{c}\right\rangle \left\vert \bar{R}R+\bar
{G}G+\bar{B}B\text{ }\right\rangle \text{ ,}%
\end{equation}
and so on and so forth.

Beyond conventional mesons, many other mesonic states are expected to exist,
most notably glueballs, which emerge as bound states of gluons.

It is interesting to notice that quantum numbers such as $J^{PC}=0^{+-},$
$J^{PC}=1^{-+},$ $J^{PC}=2^{+-},...$ \emph{cannot} be obtained in a
quark-antiquark system, but is possible for unconventional mesonic states. The
experimental discovery of mesons with these\emph{\ exotic quantum numbers
}naturally points to a non-quarkonium inner structure. Indeed, glueballs (but
also other non-conventional configurations) can produce exotic quantum numbers.

\subsection{Large-$N_{c}$}

The bare coupling constant $g_{0}$ of the QCD Lagrangian of Eq. (\ref{lqcd})
becomes, upon renormalization, a running coupling constant, which we rewrite
as:%
\begin{equation}
g^{2}(\mu)\propto\frac{1}{N_{c}\log\frac{\mu}{\Lambda_{YM}}}%
\end{equation}
Theoretically, it is very advantageous to study the limit in which $N_{c}$ is
large, since many simplifications take place. A consistent way to take the
large-$N_{c}$ limit is to postulate that $\Lambda_{YM}\propto N_{c}^{0},$ out
of which it follows that $g\propto1/\sqrt{N_{c}}.$ The following properties of
hadrons hold (see e.g. Refs \cite{thooft,witten}):

\begin{itemize}
\item The masses of quark-antiquark states and glueballs are constant for
$N_{c}\rightarrow\infty$:
\begin{equation}
M_{\left\vert \overline{q}q\right\rangle }\propto N_{c}^{0}\text{ ,
}M_{\left\vert gg\right\rangle }\propto N_{c}^{0}\text{ .}%
\end{equation}

\item The interaction between $n$ quark-antiquark states $\left\vert
\overline{q}q\right\rangle $ scales as
\begin{equation}
A_{n-\left\vert \overline{q}q\right\rangle }\propto N_{c}^{-\frac{n-2}{2}%
}\text{ }(n\geq2)\text{ .}%
\end{equation}
This implies that the amplitude for a $n$-meson scattering process becomes
smaller and smaller for increasing $N_{c}$. In particular the decay amplitude
is realized for $n=3,$ ergo $A_{\text{decay}}\propto N_{c}^{-1/2}$, implying
that the width scales as $\Gamma\propto1/N_{c}$. Conventional quarkonia become
very narrow for large $N_{c}.$

\item The interaction amplitude between $n$ glueballs $\left\vert
gg\right\rangle $ is
\begin{equation}
A_{n-\left\vert gg\right\rangle }\propto N_{c}^{-(n-2)}\text{ ,}%
\end{equation}
which is even smaller than within quarkonia.

\item The interaction amplitude between $n$ quarkonia and $m$ glueballs
behaves as
\begin{equation}
A_{\left(  n-\left\vert \overline{q}q\right\rangle \right)  \left(
m-\left\vert gg\right\rangle \right)  }\propto N_{c}^{-(\frac{n}{2}%
+m-1)}\text{ for }n\geq1\text{ and }m\geq1\text{ ,}%
\end{equation}
thus the mixing ($n=m=1$) scales as $A_{\text{mixing}}\propto N_{c}^{-1/2}.$
Then, also the glueball-quarkonium mixing is suppressed for $N_{c}\gg1$.

\item Four-quark states (both as molecular objects and diquark-antidiquark
objects). A part from a peculiar tetraquark \cite{weinberg}, these objects
typically do not survive in the large-$N_{c}$ limit.

\item Even if not relevant in this work, we recall that baryons are made of
$N_{c}$ quarks for an arbitrary $N_{c}.$ As a consequence
\begin{equation}
M_{B}\propto N_{c}\text{ .}%
\end{equation}

\end{itemize}

Indeed, the large-$N_{c}$ limit is a firm theoretical method which explains
why the quark model works. In fact, a decay channel for a certain meson causes
quantum fluctuations: the propagator of the meson is dressed by loops of other
mesons. For instance, the state $\rho^{+}$ decays into $\pi^{+}\pi^{0},$ thus
the $\rho$-meson is dressed by loops of pions. In the end, one has
schematically that the wave function of the $\rho$-meson is given by:%
\begin{equation}
\left\vert \rho^{+}\right\rangle =a\left\vert u\bar{d}\right\rangle
+b\left\vert \pi^{+}\pi^{0}\right\rangle +...\text{ ,}%
\end{equation}
where the full expression of $\left\vert u\bar{d}\right\rangle $ is given in
Eq. (\ref{rho}). Being $a\propto N_{c}^{0}$ and $b\propto N_{c}^{-1/2},$ we
understand why the quark-antiquark configuration dominates. Dots refer to
further contributions which are even more suppressed.

Yet, for $N_{c}=3$ there are some mesons for which the meson-meson component
dominates. These are for instance, the light scalar mesons that we will study
in Sec. 5.1.

\section{The construction of an effective model of low-energy QCD}

\subsection{General considerations}

QCD cannot be solved analytically. Therefore, the use of effective approaches
both in the vacuum and at finite temperature and densities represents a very
useful tool toward the understanding of QCD.

The question is the following: starting from the QCD Lagrangian $\mathcal{L}%
_{QCD}$ in terms of quarks and gluons, how should one construct the effective
Lagrangian $\mathcal{L}_{had}$ in terms of hadronic d.o.f.? Symbolically:%

\begin{equation}
\mathcal{L}_{QCD}\overset{?}{\rightarrow}\mathcal{L}_{had}\text{ .} \label{?}%
\end{equation}
There is no general accepted procedure: the hadronic Lagrangian $\mathcal{L}%
_{had}(E_{\max},N_{c}=3)$ valid up to an energy of about $E_{\max}\simeq2$ GeV
cannot be analytically obtained out of $\mathcal{L}_{QCD}$. However, the use
of symmetries turns out to be very useful, since it strongly constrains the
form that $\mathcal{L}_{had}(E_{\max},N_{c}=3)$ can have. Yet, some coupling
constants entering the Lagrangian are free parameters. (Notice that, upon
setting $E_{\max}\simeq0.6$ GeV and using the nonlinear realization of chiral
symmetry, the effective hadronic Lagrangian reduces to chiral perturbation
theory, in which only pions enter \cite{chpt}).

In the large-$N_{c}$ limit the hadronic theory must take a simple form, since
it consists of free quark-antiquark fields $\phi_{k}$ and glueball fields
$G_{h}$ (with a mass lower than $E_{\max}$):%
\begin{align}
\mathcal{L}_{had}(E_{\max},N_{c}  &  >>1,)=\sum_{k=1}^{N_{\overline{q}q}%
}\left[  \frac{1}{2}\left(  \partial_{\mu}\phi_{k}\right)  ^{2}-\frac{1}%
{2}M_{_{\overline{q}q},k}^{2}\phi_{k}^{2}\right] \nonumber\\
&  +\sum_{h=1}^{N_{gg}}\left[  \frac{1}{2}\left(  \partial_{\mu}G_{h}\right)
^{2}-\frac{1}{2}M_{G,h}^{2}G_{h}^{2}\right]  \text{ .}%
\end{align}
(Note, baryons do not appear since for $N_{c}\gg1$ their mass is very large).
From the PDG \cite{pdg} we know that below the energy $E_{\max}\simeq2$ GeV
there are various mesons with quantum numbers $J^{PC}=0^{-+},0^{++}%
,1^{--},1^{++}$ (pseudoscalar, scalar, vector, axial-vector, respectively).
But, for $N_{c}=3$ the interactions are definitely not negligible. Then, the
effective Lagrangian must include interaction terms:%
\[
\mathcal{L}_{had}(E_{\max},N_{c}=3)=
\]%
\[
\sum_{k=1}^{N_{\overline{q}q}}\left[  \frac{1}{2}\left(  \partial_{\mu}%
\phi_{k}\right)  ^{2}-\frac{1}{2}M_{_{\overline{q}q},k}^{2}\phi_{k}%
^{2}\right]  +\sum_{h=1}^{N_{gg}}\left[  \frac{1}{2}\left(  \partial_{\mu
}G_{h}\right)  ^{2}-\frac{1}{2}M_{G,h}^{2}G_{h}^{2}\right]
\]%
\begin{equation}
+\mathcal{L}_{int}^{\overline{q}q\text{-}gg}(E_{\max},N_{c}=3)+\mathcal{L}%
_{\text{new-mes}}(E_{\max},N_{c}=3)+\mathcal{L}_{\text{bar}}(E_{\max}%
,N_{c}=3)\text{ ,} \label{lhad}%
\end{equation}
where:

(i) $\mathcal{L}_{int}^{\overline{q}q\text{-}gg}(E_{\max},N_{c}=3)$ describes
the interactions of quark-antiquark with each other and with glueballs. In the
large-$N_{c}$ limit: $\mathcal{L}_{int}^{\overline{q}q\text{-}gg}(E_{\max
},N_{c})\propto O(1/N_{c}).$

(ii) $\mathcal{L}_{\text{new-mes}}(E_{\max},N_{c}=3,)$ contains (eventual!)
additional mesons, such as tetraquarks. This term disappears faster than
$O(1/N_{c})$ for large $N_{c}$.

(iii) $\mathcal{L}_{\text{bar}}(E_{\max},N_{c}=3)$ describes the baryons and
their interaction with each other and mesons.

\bigskip

In addition to the fields in the Lagrangian, it is also possible that
molecular states arise upon meson-meson interaction (see the detailed
discussion in Ref. \cite{dynrec}). These states need not to be taken into the
Lagrangian in order to avoid overcounting.

\subsection{Toward the eLSM}

An effective low-energy model of QCD should fulfill (as many as possible of)
its symmetries. Then, the properties (i)-(vi) listed in the previous section
should be present in an affective Lagrangian. Here we aim to construct an
effective model of QCD which contains from the very beginning quark-antiquark
fields with $J^{PC}=0^{-+},0^{++},1^{--},1^{++}$ as well as a scalar glueball,
which is of crucial importance for the construction of the model. We thus show
step by step the previous symmetries from the perspective of an hadronic
model. In this work we concentrate on mesons; for baryons, see Refs.
\cite{susanna} and refs. therein.

The very first observation has to do with color. Because of confinement, we
wok from the very beginning with white objects, denoted as $\Phi,R_{\mu},...,$
see below. Thus, they are obviously invariant under $SU\left(  3\right)  _{c}$
: $\Phi\rightarrow\Phi,$ $R_{\mu}\rightarrow R_{\mu},...$ invariance under
$SU(N_{c}=3)_{c}$ is automatically fulfilled. Yet, the parameters of the model
will depend on the number of colors $N_{c}$.

\subsubsection{\textit{Yang-Mills sector}}

We need to describe the trace anomaly correctly. For this reason, we first
concentrate on the pure YM sector. We introduce an effective field $G$ which
`intuitively' corresponds to
\begin{equation}
G^{4}\sim G_{\mu\nu}^{a}G^{a,\mu\nu}\text{ .}%
\end{equation}
Then, $G$ is a collective field describing gluons. As shown in Ref.
\cite{migdal,salo,ellis} the following Lagrangian
\begin{equation}
\mathcal{L}_{dil}=\frac{1}{2}(\partial_{\mu}G)^{2}-V_{dil}(G)\text{ , }%
\end{equation}
with%
\begin{equation}
V_{dil}(G)=\frac{1}{4}\frac{m_{G}^{2}}{\Lambda_{G}^{2}}\left[  G^{4}\ln\left(
\frac{G}{\Lambda_{G}}\right)  -\frac{G^{4}}{4}\right]  \label{vdil}%
\end{equation}
is what we need. Because of the logarithm and the dimensional parameter
$\Lambda_{G},$ dilatation symmetry
\[
x^{\mu}\rightarrow\lambda^{-1}x^{\mu}\text{ and }G(x)\rightarrow G^{\prime
}(x^{\prime})=\lambda G(x)\text{ }%
\]
is explicitly broken. In fact, the divergence of the associated Noether
current is:
\begin{equation}
\partial_{\mu}J^{\mu}=T_{\mu}^{\mu}=G\partial_{G}V_{dil}(G)-4G=-\frac{1}%
{4}\frac{m_{G}^{2}}{\Lambda_{G}^{2}}G^{4}\text{ .}%
\end{equation}
There is a clear correspondence to Eq. (\ref{sa}). Upon taking the v.e.v. we
have
\begin{equation}
\left\langle T_{\mu}^{\mu}\right\rangle =\left\langle -\frac{1}{4}\frac
{m_{G}^{2}}{\Lambda_{G}^{2}}G^{4}\right\rangle =-\frac{1}{4}\frac{m_{G}^{2}%
}{\Lambda_{G}^{2}}G_{0}^{4}\equiv-\left\langle \frac{11N_{c}}{48}\frac
{\alpha_{s}}{\pi}G_{\mu\nu}^{a}G^{a,\mu\nu}\right\rangle \text{ .}%
\end{equation}
It is easy to prove that $G_{0}=\Lambda_{G}$ corresponds to the minimum of
$V_{dil}(G).$ Thus, the emergence of a gluon condensate can be easily
understood in this effective framework.

By studying the fluctuations about the minimum, $G\rightarrow G_{0}+G$, one
can see that a field with mass $m_{G}$ emerges. This particle is the famous
scalar glueball. This is, according to lattice simulations
\cite{mainlattice,morningstar}, the lightest glueball with $m_{G}\sim
1.6$-$1.7$ GeV. As we shall see, $f_{0}(1710)$ is a very good candidate.

The lightest scalar glueball is very important since it is related to
dilatation symmetry, but further gluonic fields can be easily introduced. For
illustrative purposes, by restricting to scalar and pseudoscalar glueballs, we
have:
\begin{equation}
\mathcal{L}=\mathcal{L}_{dil}+\frac{1}{2}\sum_{k}\alpha_{k}G_{k}^{2}G^{2}%
+\sum_{k,l}\beta_{k,l}G_{k}^{2}G_{l}^{2}\text{ ,} \label{glueballgen}%
\end{equation}
where $\alpha_{k}$ is the interaction with the lightest glueball $G$ and
$\beta_{k,l}$ further interactions. The mas of the $k$-glueball emerges upon
condensation of $G$: $M_{G_{k}}^{2}=\alpha_{k}G_{0}^{2}$

As a last step, we describe the large-$N_{c}$ dependence of the parameters.
The mass $m_{G}$ is independent on $N_{c}$, while $\Lambda_{G}$ scales as
$N_{c},$ in such a way that $G^{4}$-interaction scales as $1/N_{c}^{2}$ :%
\begin{equation}
m_{G}\propto N_{c}^{0}\text{ , }\Lambda_{G}\propto N_{c}\text{ .}%
\end{equation}
The parameters $\alpha_{k}$ and $\beta_{k,l}$ in Eq. (\ref{glueballgen}) scale
as $1/N_{c}^{2}$.

Next, we leave the gluonic sector and turn our attention to quark-antiquark fields.

\subsubsection{Scalar and pseudoscalar mesons}

Scalar and pseudoscalar quark-antiquark mesons are contained in the
$N_{f}\times N_{f}$ matrix $\Phi$, which corresponds to the following
quark-antiquark substructure:
\begin{equation}
\Phi_{ij}\equiv\sqrt{2}\overline{q}_{j,R}q_{i,L}\text{ .}%
\end{equation}
The equivalence $\equiv$ means that $\Phi$ and $\sqrt{2}\overline{q}%
_{j,R}q_{i,L}$ transform in the same way under chiral transformation, but it
does not mean that $\Phi$ is a perturbative quark-antiquark object.
Intuitively, the quark-antiquark current is dressed by gluonic clouds. Then,
$\Phi$ is an effective object in which constituent quarks enter (a more
detailed correspondence can be realized by using nonlocal currents, see e.g.
Refs. \cite{lyubo,terning,glueball1}). Yet, the identification $\Phi
_{ij}\equiv\sqrt{2}\overline{q}_{j,R}q_{i,L}$ is sufficient to study
transformation properties.

We recall that upon chiral transformation $q_{i,L}\rightarrow U_{L}q_{i,L},$
$q_{i,R}\rightarrow U_{R}q_{i,R},$ then the field $\Phi$ transforms as:%
\begin{equation}
\Phi\rightarrow U_{L}\Phi U_{R}^{\dagger}\text{ .}%
\end{equation}
[Under $U_{V}(1)$ on has $U_{L}=U_{R}=e^{i\theta t^{0}}$, then: $\Phi
\rightarrow\Phi$]. $\Phi$ can also be expressed as:
\begin{align}
\Phi_{ij} &  \equiv\sqrt{2}\overline{q}_{j,R}q_{i,L}=\sqrt{2}\overline{q}%
_{j}P_{L}P_{L}q_{i}=\sqrt{2}\overline{q}_{j}P_{L}q_{i}\nonumber\\
&  =\frac{1}{\sqrt{2}}\left(  \overline{q}_{j}q_{i}-\overline{q}_{j}\gamma
^{5}q_{i}\right)  =\frac{1}{\sqrt{2}}\left(  \overline{q}_{j}q_{i}%
+i\overline{q}_{j}i\gamma^{5}q_{i}\right)
\end{align}
where $P_{L}=\frac{1}{2}(1-\gamma^{5})$ and $P_{R}=\frac{1}{2}(1+\gamma^{5})$
are the chiral projectors. One recognizes scalar and pseudoscalar objects:
\begin{equation}
S_{ij}\equiv\overline{q}_{j}q_{i}\text{ , }\mathcal{P}_{ij}\equiv\overline
{q}_{j}i\gamma^{5}q_{i}\text{ ,}%
\end{equation}
and finally:
\begin{equation}
\Phi=\mathcal{S}+i\mathcal{P}\text{ .}%
\end{equation}
The matrices $\mathcal{S}$ and $\mathcal{P}$ are Hermitian, so they can be
written as:%
\begin{equation}
\mathcal{S}=S^{a}t^{a}\text{ , }\mathcal{P}=P^{a}t^{a}\text{ ,}%
\end{equation}%
\begin{equation}
S^{a}\equiv\sqrt{2}\overline{q}t^{a}q\text{ , }P^{a}\equiv\sqrt{2}\overline
{q}i\gamma^{5}t^{a}q\text{ .}\label{sp}%
\end{equation}
The transformation properties of $\Phi,$ $\mathcal{S}$, and $\mathcal{P}$ are
summarized in Table 1.

\bigskip

\bigskip

\begin{center}
\textbf{Table 1}: Transformation of $\mathcal{P},$ $\mathcal{S}$ and $\Phi$
(from \cite{hab}).%

\begin{tabular}
[c]{|c|c|c|c|}\hline
& $\mathcal{P}$ & $\mathcal{S}$ & $\Phi$\\\hline
Elements & $\mathcal{P}_{ij}\equiv\overline{q}_{j}i\gamma^{5}q_{i}$ &
$\mathcal{S}_{ij}\equiv\overline{q}_{j}q_{i}$ & $\Phi_{ij}\equiv\sqrt
{2}\overline{q}_{j,R}q_{i,L}$\\\hline
Currents & $P^{i}\equiv\overline{q}i\gamma^{5}\frac{\lambda_{i}}{\sqrt{2}}q$ &
$S^{i}\equiv\overline{q}\frac{\lambda_{i}}{\sqrt{2}}q$ & $\Phi^{i}\equiv
\sqrt{2}\overline{q}_{R}\frac{\lambda_{i}}{\sqrt{2}}q_{L}$\\\hline
P & $-\mathcal{P(}x^{0},-\mathbf{x}\mathcal{)}$ & $\mathcal{S(}x^{0}%
,-\mathbf{x}\mathcal{)}$ & $\Phi^{\dagger}\mathcal{(}x^{0},-\mathbf{x}%
\mathcal{)}$\\\hline
C & $\mathcal{P}^{t}$ & $\mathcal{S}^{t}$ & $\Phi^{t}$\\\hline
$U(N_{f})_{V}$ & $U_{V}\mathcal{P}U_{V}^{\dagger}$ & $U_{V}\mathcal{S}%
U_{V}^{\dagger}$ & $U_{V}\Phi U_{V}^{\dagger}$\\\hline
$U(N_{f})_{A}$ & $\frac{1}{2i}\left(  U_{A}\Phi U_{A}-U_{A}^{\dagger}%
\Phi^{\dagger}U_{A}^{\dagger}\right)  $ & $\frac{1}{2}\left(  U_{A}\Phi
U_{A}+U_{A}^{\dagger}\Phi^{\dagger}U_{A}^{\dagger}\right)  $ & $U_{A}\Phi
U_{A}$\\\hline
$U(N_{f})_{R}\times$ $U(N_{f})_{L}$ & $\frac{1}{2i}\left(  U_{L}\Phi
U_{R}^{\dagger}-U_{R}\Phi^{\dagger}U_{L}^{\dagger}\right)  $ & $\frac{1}%
{2}\left(  U_{L}\Phi U_{R}^{\dagger}+U_{R}\Phi^{\dagger}U_{L}^{\dagger
}\right)  $ & $U_{L}\Phi U_{R}^{\dagger}$\\\hline
\end{tabular}

\bigskip
\end{center}

In the chiral limit $m_{i}\rightarrow0$ chiral symmetry is exact. The
effective Lagrangian $\mathcal{L}_{\Phi}$ including only the field $\Phi$
reads (the $U_{A}(1)$ anomaly is neglected, see later):
\begin{equation}
\mathcal{L}_{\Phi}=\mathrm{Tr}\left[  (\partial^{\mu}\Phi)^{\dagger}%
(\partial_{\mu}\Phi)-m_{0}^{2}\Phi^{\dagger}\Phi-\lambda_{2}\left(
\Phi^{\dagger}\Phi\right)  ^{2}\right]  -\lambda_{1}(\mathrm{Tr}[\Phi
^{\dagger}\Phi])^{2}\text{ .}\label{lsigma}%
\end{equation}
This is the famous $\sigma$-model of QCD with (pseudo)scalar quarkonia.
Historically, it has been a very important tool to study chiral symmetry and
its breaking. In the present form it is not realistic enough, since the
results for the decay turn out not to be consistent. A realistic version of
the $\sigma$-model is obtained by including (axial-)vector d.o.f., as we
discuss in Sec. 3.3.

\subsubsection{The simple case $N_{f}=1$}

For pedagogical purposes, it is very instructive to study the case $N_{f}=1,$
in which only one $\sigma$ and only one pion $\pi$ are present: $\Phi
=\mathcal{\sigma}+i\mathcal{\pi}$ (see Ref. \cite{nikolla}). In this case,
chiral symmetry is a simple rotation in the $(\pi,\sigma)$-plane corresponding
to $U(N_{f}=1)_{A}.$ (In reality, this symmetry is broken because of the axial
anomaly. However, here we do not consider the anomaly but we simply regard
$U(1)_{A}$ as a simplified limiting case of chiral symmetry):
\begin{equation}
\left(
\begin{array}
[c]{c}%
\pi\\
\sigma
\end{array}
\right)  \rightarrow\left(
\begin{array}
[c]{cc}%
\cos\theta & \sin\theta\\
-\sin\theta & \cos\theta
\end{array}
\right)  \left(
\begin{array}
[c]{c}%
\pi\\
\sigma
\end{array}
\right)  \text{ .}%
\end{equation}
The potential in\ Eq. (\ref{lsigma}) takes the simple form:%
\begin{equation}
V(\sigma,\pi)=\frac{m_{0}^{2}}{2}\left(  \pi^{2}+\sigma^{2}\right)
+\frac{\lambda}{4}\left(  \pi^{2}+\sigma^{2}\right)  ^{2}\text{ ,}%
\end{equation}
which is manifestly chirally invariant (since it depends on $\pi^{2}%
+\sigma^{2}$ only). For $m_{0}^{2}>0$ the potential has a single minimum for
$P_{\min}=(\sigma=0,\pi=0),$ see Fig.1.

\begin{figure}[tbh]
\centering
\includegraphics[width=0.50\textwidth]{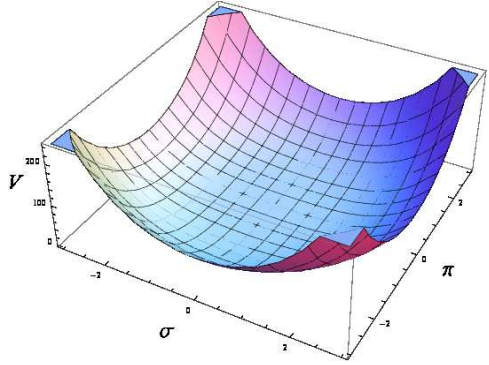} \caption{Form of the
potential for $m_{0}^{2}>0$. A single minimum located in the origin is
present.}%
\label{fig1}%
\end{figure}

In fact, along the $\sigma$-direction:%
\begin{equation}
\partial_{\sigma}V(\sigma,0)=m_{0}^{2}\sigma+\lambda\sigma^{3}=0\text{ , }%
\end{equation}
out of which
\begin{equation}
\sigma=0\text{ and }\sigma=\pm\sqrt{\frac{-m_{0}^{2}}{\lambda}}\text{ .}%
\end{equation}
For $m_{0}^{2}>0$ the value $\sigma=0$ is the only solution. The masses of the
particles corresponds to the second derivatives evaluated at the minimum:%
\begin{align}
m_{\pi}^{2}  &  =\left.  \frac{\partial^{2}V}{\partial\pi^{2}}\right\vert
_{P=P_{\min}}=m_{0}^{2}\text{ ,}\\
m_{\sigma}^{2}  &  =\left.  \frac{\partial^{2}V}{\partial\sigma^{2}%
}\right\vert _{P=P_{\min}}=m_{0}^{2}\text{ .}%
\end{align}
As expected, both particles have the same mass $m_{0}.$ This is a direct
consequence of chiral symmetry.

We know however that Nature is \textit{not} like that. The pion is nearly
massless ($\simeq135$ MeV) but the sigma field, which shall be identified with
$f_{0}(1370),$ is much heavier: $1350\pm150$ MeV. The reason for that is the
spontaneous symmetry breaking. We can describe it in our model by considering
\begin{equation}
m_{0}^{2}<0\text{ . }%
\end{equation}
(Then, $m_{0}$ is purely imaginary). The corresponding potential can be
rewritten as:
\begin{equation}
V(\sigma,\pi)=\frac{\lambda}{4}\left(  \pi^{2}+\sigma^{2}-F^{2}\right)
^{2}+const\text{ .}%
\end{equation}
It has the typical shape of a Mexican hat, in which the origin is not a
minimum but a maximum, see Fig. 2.

\begin{figure}[tbh]
\centering
\includegraphics[width=0.50\textwidth]{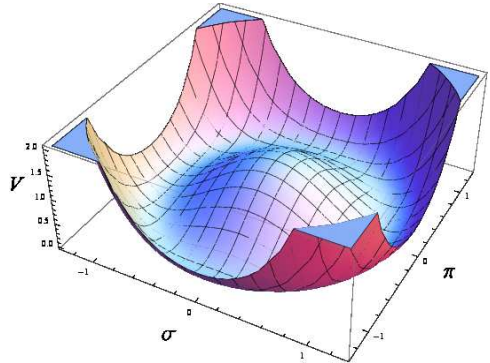} \caption{Form of the potential
for $m_{0}^{2}<0$. A circle of minima is present.}%
\label{fig2}%
\end{figure}

The origin is not stable. In fact, upon calculating the masses around the
origin, they would turn out to be imaginary. There is however a circle of
equivalent minima for:
\begin{equation}
\pi^{2}+\sigma^{2}=-\frac{m_{0}^{2}}{\lambda}>0\text{ .}%
\end{equation}

According to spontaneous symmetry breaking, only one of them is realized.
Clearly we cannot predict which one. We choose (they are all equivalent):%

\begin{equation}
P_{\min}=\left(  F=\sqrt{-\frac{m_{0}^{2}}{\lambda}},0\right)  \text{ .}%
\end{equation}
After having made this choice, the system has undergone spontaneous chiral
symmetry breaking. The value of the $\sigma$-field at the minimum is given
by:
\begin{equation}
\sigma_{\min}=\phi_{N}=\sqrt{-\frac{m_{0}^{2}}{\lambda}}=F\text{ .}%
\end{equation}
$\phi_{N}$ is also denoted as the chiral condensate and is indeed proportional
to $\left\langle 0_{QCD}\left\vert \bar{q}q\right\vert 0_{QCD}\right\rangle
\neq0.$

We now calculate the masses:%
\begin{align}
m_{\pi}^{2}  &  =\left.  \frac{\partial^{2}V}{\partial\pi^{2}}\right\vert
_{P=P_{\min}}=0\text{ ,}\\
m_{\sigma}^{2}  &  =\left.  \frac{\partial^{2}V}{\partial\sigma^{2}%
}\right\vert _{P=P_{\min}}=m_{0}^{2}+3\lambda\phi_{N}^{2}=-2m_{0}^{2}%
=2\lambda\phi_{N}^{2}>0\text{ ,}%
\end{align}
which are now very different from each other: the pion is massless as a
consequence of the Goldstone theorem. On the contrary, the mass of the
$\sigma$ is nonzero. One then realizes how spontaneous chiral symmetry
breaking generates different masses for chiral partners. Note, $m_{\sigma}$ is
proportional to the chiral condensate $\phi_{N}.$

In Nature, the pion is not exactly masses. In order to describe this fact, we
modify the potential as%

\begin{equation}
V(\sigma,\pi)=\frac{m_{0}^{2}}{2}\left(  \pi^{2}+\sigma^{2}\right)
+\frac{\lambda}{4}\left(  \pi^{2}+\sigma^{2}\right)  ^{2}-h\sigma\text{ ,}%
\end{equation}
where $-h\sigma$ breaks chiral symmetry explicitly. This term follows directly
from the mass term $-m\bar{q}q$ in the QCD Lagrangian. We thus expect that
$h\propto m,$ where $m$ is the bare quark mass (for instance, the $u$ quark or
the average $(m_{u}+m_{d})/2$). The form of the potential is depicted in Fig. 3.

\bigskip

\begin{figure}[tbh]
\centering
\includegraphics[width=0.50\textwidth]{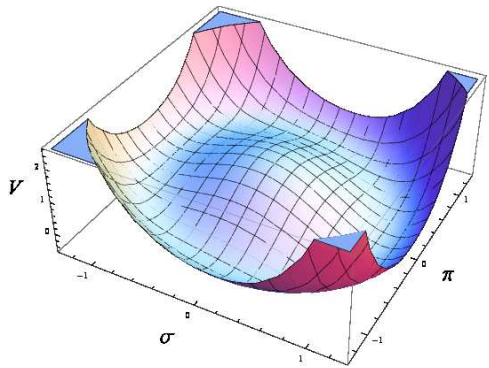} \caption{Form of the potential
in presence of explicit symmetry breaking.}%
\label{fig3}%
\end{figure}

\bigskip

Now, it has a unique minimum along the $\sigma$ direction. The corresponding
equation for $\phi_{N}$ reads:%
\begin{align}
\left.  \frac{\partial V(\sigma,0)}{\partial\sigma}\right\vert _{\sigma
=\sigma_{\min}=\phi_{N}}  &  =m_{0}^{2}\phi_{N}+\lambda\phi_{N}^{3}-h\text{
}\nonumber\\
&  =\phi_{N}\left(  m_{0}^{2}+\lambda\phi_{N}^{2}\right)  -h=0\text{ ,}%
\end{align}
which is of third order. Only one of the three equation is physical, see Ref.
\cite{denisdiss}. The minimum is denoted as $P_{\min}=(\phi_{N},0).$ The pion
mass is now nonzero:
\begin{equation}
m_{\pi}^{2}=\left.  \frac{\partial^{2}V}{\partial\pi^{2}}\right\vert
_{P=P_{\min}}=m_{0}^{2}+\lambda\phi_{N}^{2}=\frac{h}{\phi_{N}}>0\text{ .}%
\end{equation}
We realize that the pion mass scale as $m_{\pi}\propto\sqrt{h}\propto\sqrt
{m}.$ This nontrivial dependence is a signature of an explicit symmetry
breaking on top of a spontaneous symmetry breaking. For $h\rightarrow0$ the
pion mass vanishes, as expected.

The mass of the $\sigma$-particle is:%

\begin{equation}
m_{\sigma}^{2}=\left.  \frac{\partial^{2}V}{\partial\sigma^{2}}\right\vert
_{P=P_{\min}}=m_{0}^{2}+3\lambda\phi_{N}^{2}=m_{\pi}^{2}+2\lambda\phi_{N}%
^{2}\text{ .}%
\end{equation}
Again, the mass of the $\sigma$-particle is always larger than the pion mass,
The difference $m_{\sigma}^{2}-m_{\pi}^{2}=2\lambda\phi_{N}^{2}>0$ is not
explicitly dependent on $h.$

The chiral condensate $\phi_{N}$ affects the masses and also decays, such as
$\sigma\rightarrow\pi\pi$. The latter is determined in the following way: one
first performs the shift $\sigma\rightarrow\sigma+\phi_{N}$ and then expands
the potential. The term causing the decay is given by $\lambda\phi_{N}%
\sigma\pi^{2}.$ Hence:%
\begin{equation}
\Gamma_{\sigma\rightarrow\pi^{2}}=s\frac{\left\vert \vec{k}\right\vert }{8\pi
m_{\sigma}^{2}}\left[  \lambda\phi_{N}\right]  ^{2}\text{ ,}%
\end{equation}
where $\left\vert \vec{k}\right\vert =\sqrt{\frac{m_{\sigma}^{2}}{4}-m_{\pi
}^{2}}$ is the modulus of the three-momentum of one of the outgoing particles,
and $s$ is a symmetry and isospin factor ($s=2$ in the present version of the
model with only one type of pion, but $s=6$ when three pions are taken into account).

One can also show that $\phi_{N}$ enters into the expressions of the weak
decay of pions. In the present version of the model, the condensate
corresponds to the pion-decay constant: $\phi_{N}=f_{\pi}=92.4$ MeV
\cite{koch}.

In the full $N_{f}=3$ version of the model, the masses and decays are
calculated by following the very same steps. Obviously, there are much more
fields and decay channels, but the principle and the basic ideas are exactly
the same as those discussed here.

\subsubsection{A criterion for the construction of an hadronic model}

We now come back to the Lagrangian of Eq. (\ref{lsigma}). One may wonder why
terms of the type $(Tr[\Phi^{+}\Phi])^{6},$ which are also chirally symmetric,
were not included. The fact that such a term would break renormalization is
not a good argument: an effective hadronic theory is valid only in the
low-energy sector and does not need to be renormalizable.

In order to understand this point, we need to introduce the dilaton field $G$.
The corresponding Lagrangian has the general form:
\begin{equation}
\mathcal{L}_{G\Phi}=\frac{1}{2}(\partial_{\mu}G)^{2}-V_{dil}(G)+\mathrm{Tr}%
\left[  (\partial^{\mu}\Phi)^{\dagger}(\partial^{\mu}\Phi)\right]  -V_{G\Phi
}\mathcal{(}G,\Phi)\text{ ,}%
\end{equation}
where $V_{dil}(G)=\frac{1}{4}\frac{m_{G}^{2}}{\Lambda_{G}^{2}}\left[  G^{4}%
\ln\left(  \frac{G}{\Lambda_{G}}\right)  -\frac{G^{4}}{4}\right]  $ was
introduced in Eq. (\ref{vdil}). Now, we require the two following properties:

(1) In the chiral limit $(m_{i}=0)$ there is in $\mathcal{L}_{G\Phi}$ a single
dimensional parameter: the scale $\Lambda_{G}$ entering in the potential
$V_{dil}(G)$. In this way the trace-anomaly is generated in the YM sector, in
accordance with QCD.

(2) The potential $V_{G\Phi}\mathcal{(}G,\Phi)$ is finite when the fields are
finite (smoothness of the potential).

These two criteria severely constrain the form of $\mathcal{L}_{G\Phi}.\ $In
fact, a term of the type
\begin{equation}
\alpha(\mathrm{Tr}[\Phi^{\dagger}\Phi])^{6}\text{ ,}%
\end{equation}
is not allowed because of (1) (the parameter $\alpha$ has dimension
Energy$^{-2}$). One could still modify the term as%
\begin{equation}
\frac{\beta}{G^{2}}(\mathrm{Tr}[\Phi^{\dagger}\Phi])^{6}%
\end{equation}
in which the constant $\beta$ is dimensionless and thus in agreement with (1).
But point (2) is broken, since this term is singular for $G=0$. The validity
of point (2) assures also that there is a smooth limit at high temperatures in
which the gluon condensate $G_{0}$ is expected to become small.

As a consequence of (1) and (2), one has to consider only interaction terms
with dimension \textit{exactly} equal to four:%
\begin{align}
\mathcal{L}_{G\Phi}  &  =\frac{1}{2}(\partial_{\mu}G)^{2}-V_{dil}%
(G)\nonumber\\
&  +\mathrm{Tr}\left[  (\partial^{\mu}\Phi)^{\dagger}(\partial_{\mu}%
\Phi)-aG^{2}\Phi^{\dagger}\Phi-\lambda_{2}\left(  \Phi^{\dagger}\Phi\right)
^{2}\right]  -\lambda_{1}(\mathrm{Tr}[\Phi^{\dagger}\Phi])^{2}\text{ .}
\label{lgphi}%
\end{align}
The term $aG^{2}\Phi^{+}\Phi$ describes the interaction of the glueball and
(pseudo)scalar mesons. The connection to $\mathcal{L}_{\Phi}$ of Eq.
(\ref{lsigma}) is clear:
\begin{equation}
m_{0}^{2}=aG_{0}^{2}\text{ ,} \label{m0g0}%
\end{equation}
where $a$ is dimensionless. (In presence of quarkonia $G_{0}\simeq$
$\Lambda_{G}$ but is not exactly equal because mixing arises.) In conclusion,
it is not renormalization but scale invariance (broken solely by $\Lambda
_{G})$ which obliges us to take into account terms with dimension exactly
equal to 4.

The large-$N_{c}$ dependence of the parameters reads:%
\begin{align}
a  &  \propto N_{c}^{-2}\rightarrow m_{0}^{2}\propto N_{c}^{0}\text{ ,}\\
\lambda_{2}  &  \propto N_{c}^{-1}\text{ ,}\\
\lambda_{1}  &  \propto N_{c}^{-2}\text{ .}%
\end{align}
The $\lambda_{2}$-term scales as $N_{c}^{-1}$ in agreement with Section 2.3.
However, $\lambda_{1}$ scales as $N_{c}^{-2}$ because it arises as the product
of two traces. One can have transitions in which all quark lines are
annihilated, what implies an additional suppression of $N_{c}.$

\subsubsection{Spontaneous symmetry breaking revisited}

We already described spontaneous symmetry breaking in Sec. 3.2.3. In the
present section, we reconsider it in the presence of $G.$ Upon setting
$\Phi=\sigma t^{0}$ the potential reads:
\begin{equation}
V(G,\sigma)=V_{dil}(G)+aG^{2}\sigma^{2}+(\lambda_{2}+\lambda_{1})\sigma
^{4}\text{.} \label{vgphi}%
\end{equation}
Then:

$\bullet$ $a>0\rightarrow$ $m_{0}^{2}=aG_{0}^{2}>0.$ Minimum for $G_{0}\neq0,$
$\sigma_{0}=0$,

$\bullet$ $a<0\rightarrow$ $m_{0}^{2}=aG_{0}^{2}<0.$ Minimum for $G_{0}\neq0,$
$\sigma_{0}\neq0.$

Then, for $a<0,$ spontaneous breaking of chiral symmetry is realized for%
\begin{equation}
\sigma_{0}\sim\sqrt{-\frac{m_{0}^{2}}{\lambda_{1}+\lambda_{2}}}\sim
\sqrt{-\frac{a}{\lambda_{1}+\lambda_{2}}}G_{0}\text{ .}%
\end{equation}

Being $G_{0}\sim\Lambda_{G},$ it follows that $\sigma_{0}\propto\Lambda_{G}.$
Thus, the chiral condensate is also a consequence of the breaking of
dilatation symmetry. Indeed, the trace anomaly is a very fundamental property
of QCD which is at the basis of all low-energy hadronic phenomena.

\subsubsection{$U(1)_{A}$\emph{ }anomaly}

The $U(1)_{A}$ anomaly is taken into account by the following additional term:%
\begin{equation}
\mathcal{L}_{U(1)_{A}}=c_{1}(\det\Phi^{\dagger}-\det\Phi)^{2}\text{ .}
\label{an}%
\end{equation}
This term is invariant under $SU(N_{f})_{R}\times SU(N_{f})_{L}$. In fact,
using $\det[ABC]=\det[A]\det[B]\det[C]$ one can see that $\Phi\rightarrow
U_{L}\Phi U_{R}^{\dagger}$ does not change $\mathcal{L}_{U(1)_{A}}.$ This term
is however not invariant under $U(1)_{A}$, for which $\Phi\rightarrow e^{2i\nu
t^{0}}\Phi.$ In fact:
\begin{align}
\mathcal{L}_{U(1)_{A}}  &  =c_{1}\left(  \det\Phi^{\dag}-\det\Phi\right)
^{2}\nonumber\\
&  \rightarrow c_{1}\left(  e^{-i\nu\sqrt{2N_{f}}}\det\Phi^{\dag}-e^{i\nu
\sqrt{2N_{f}}}\det\Phi\right)  ^{2}\neq\mathcal{L}_{U(1)_{A}}\text{ .}
\label{det}%
\end{align}
This term influence the masses of the (pseudo)scalar mesons and is responsible
for the large mass of $\eta^{\prime}$ ($\sim$ 1 GeV).

\subsubsection{Vector and Axial-vector mesons}

Vector and axial-vector mesons are light ($\lesssim1.4$ GeV) and \textit{must}
be included in a realistic mesonic model of QCD. A general description of the
issue was already presented in Ref. \cite{geffen}. In Refs.
\cite{nf2,ko,buballa} chiral models for $N_{f}=2$ were explicitly constructed.
Yet, the full $N_{f}=3$ case was constructed only very recently in\ Ref.
\cite{dick}.

We now repeat the previous steps for (axial-)vector states. Mathematically,
one introduces $N_{f}\times N_{f}$ matrices $R_{\mu}$ and $L_{\mu}:$
\begin{equation}
R_{ij}^{\mu}\equiv\sqrt{2}\overline{q}_{j,R}\gamma^{\mu}\overline{q}%
_{i,R}=\frac{1}{\sqrt{2}}\left(  \overline{q}_{j}\gamma^{\mu}\overline{q}%
_{i}-\overline{q}_{j}\gamma^{5}\gamma^{\mu}\overline{q}_{i}\right)  \text{ ,}%
\end{equation}%
\begin{equation}
L_{ij}^{\mu}\equiv\sqrt{2}\overline{q}_{j,R}\gamma^{\mu}\overline{q}%
_{i,R}=\frac{1}{\sqrt{2}}\left(  \overline{q}_{j}\gamma^{\mu}\overline{q}%
_{i}+\overline{q}_{j}\gamma^{5}\gamma^{\mu}\overline{q}_{i}\right)  \text{ .}%
\end{equation}
Under chiral transformations they transform as:%
\begin{equation}
R^{\mu}\rightarrow U_{R}R^{\mu}U_{R}^{\dagger}\text{ , }L^{\mu}\rightarrow
U_{L}L^{\mu}U_{L}^{\dagger}\text{ .}%
\end{equation}
The matrices $R_{\mu}$ and $L_{\mu}$ are linear combinations of the vector and
axial-vector fields $V_{\mu}$ and $A_{\mu}$
\begin{equation}
R^{\mu}=V^{\mu}-A^{\mu}\text{ },
\end{equation}%
\begin{equation}
L^{\mu}=V^{\mu}+A^{\mu}\text{ ,}%
\end{equation}
with:
\begin{align}
V_{ij}^{\mu}  &  \equiv\frac{1}{\sqrt{2}}\overline{q}_{j}\gamma^{\mu}%
\overline{q}_{i}=V^{\mu,a}t^{a}\text{ };V^{\mu,a}\equiv\sqrt{2}\overline
{q}\gamma^{\mu}t^{a}q\text{ ,}\\
\text{ }A_{ij}^{\mu}  &  \equiv\frac{1}{\sqrt{2}}\overline{q}_{j}\gamma
^{5}\gamma^{\mu}\overline{q}_{i}=A^{\mu,a}t^{a}\text{ };A^{\mu,a}\equiv
\sqrt{2}\overline{q}\gamma^{5}\gamma^{\mu}t^{a}q\text{ .}%
\end{align}
Tables 2 and 3 show the transformations of $R^{\mu},L^{\mu}$ and $V_{\mu}$,
$A_{\mu}.$

\bigskip

\begin{center}
\textbf{Table 2}: Transformations of $R_{\mu}$ and $L_{\mu}$ \cite{hab}%

\begin{tabular}
[c]{|c|c|c|}\hline
& $R_{\mu}$ & $L_{\mu}$\\\hline
Elements & $R_{ij}^{\mu}\equiv\sqrt{2}\overline{q}_{j,R}\gamma^{\mu}q_{i,R}$ &
$L_{ij}^{\mu}\equiv\sqrt{2}\overline{q}_{j,R}\gamma^{\mu}q_{i,R}$\\\hline
Currents & $R_{\mu}^{i}\equiv\overline{q}_{R}\gamma^{\mu}\frac{\lambda_{i}%
}{\sqrt{2}}q_{R}$ & $L_{\mu}^{i}\equiv\overline{q}_{L}\gamma^{\mu}%
\frac{\lambda_{i}}{\sqrt{2}}q_{L}$\\\hline
$P$ & $g^{\mu\nu}L_{\mu}\mathcal{(}x^{0},-\mathbf{x}\mathcal{)}$ & $g^{\mu\nu
}R_{\mu}\mathcal{(}x^{0},-\mathbf{x}\mathcal{)}$\\\hline
$C$ & $-L_{\mu}^{t}$ & $R_{\mu}^{t}$\\\hline
$U(N_{f})_{V}$ & $U_{V}R_{\mu}U_{V}^{\dagger}$ & $U_{V}L_{\mu}U_{V}^{\dagger}%
$\\\hline
$U(N_{f})_{A}$ & $U_{A}R_{\mu}U_{A}^{\dagger}$ & $U_{A}^{\dagger}L_{\mu}U_{A}%
$\\\hline
$U(N_{f})_{R}\times$ $U(N_{f})_{L}$ & $U_{R}R_{\mu}U_{R}^{\dagger}$ &
$U_{L}R_{\mu}U_{L}^{\dagger}$\\\hline
\end{tabular}

\bigskip

\bigskip

\textbf{Table 3}: Transformations of $V_{\mu}$ and $A_{\mu}$ \cite{hab}%

\begin{tabular}
[c]{|c|c|c|}\hline
& $V_{\mu}$ & $A_{\mu}$\\\hline
Elements & $V_{ij}^{\mu}\equiv\sqrt{2}\overline{q}_{j}\gamma^{\mu}q_{i}$ &
$A_{ij}^{\mu}\equiv\sqrt{2}\overline{q}_{j}\gamma^{5}\gamma^{\mu}q_{i}%
$\\\hline
Currents & $V^{i}\equiv\overline{q}\gamma^{\mu}\frac{\lambda_{i}}{\sqrt{2}}q$
& $A^{i}\equiv\overline{q}\gamma^{5}\gamma^{\mu}\frac{\lambda_{i}}{\sqrt{2}}%
q$\\\hline
$P$ & $g^{\mu\nu}V_{\mu}\mathcal{(}x^{0},-\mathbf{x}\mathcal{)}$ & $-g^{\mu
\nu}A_{\mu}\mathcal{(}x^{0},-\mathbf{x}\mathcal{)}$\\\hline
$C$ & $-V_{\mu}^{t}$ & $A_{\mu}^{t}$\\\hline
\end{tabular}

\bigskip
\end{center}

Notice that under parity a vector-field, such as the $\rho$-meson, transforms
as $\rho^{i}(t,\mathbf{x})\rightarrow-\rho^{i}(t,-\mathbf{x})$, $\rho
^{0}(t,\mathbf{x})\rightarrow\rho^{0}(t,-\mathbf{x})$. This is why at nonzero
density the field $\omega^{0}$ and $\rho^{0}$ (may) condense.

The corresponding Lagrangian $\mathcal{L}_{AV}$ is constructed by following
the very same principles of Sec. 3.2.1 and 3.2.4. It consists of the sum
\begin{equation}
\mathcal{L}_{AV}=\mathcal{L}_{2,AV}+\mathcal{L}_{3,AV}+\mathcal{L}%
_{4,AV}+\mathcal{L}_{PS,AV}\text{ ,} \label{lav}%
\end{equation}
where 2, 3, 4 is the number of (axial-)vector fields at each vertex and where
$\mathcal{L}_{\Phi,AV}$ describes the interaction with (pseudo)scalar
quarkonium states.

The term $\mathcal{L}_{2,AV}$ reads
\begin{equation}
\mathcal{L}_{2,AV}=-\frac{1}{4}\mathrm{Tr}[(L^{\mu\nu})^{2}+(R^{\mu\nu}%
)^{2}]+\frac{b}{2}G^{2}\mathrm{Tr}[(L^{\mu})^{2}+(R^{\mu})^{2}]\text{ ,}
\label{l2av}%
\end{equation}%
\begin{equation}
L^{\mu\nu}=\partial^{\mu}L^{\nu}-\partial^{\nu}L^{\mu},\text{ }R^{\mu\nu
}=\partial^{\mu}R^{\nu}-\partial^{\nu}R^{\mu}\text{ .}%
\end{equation}
When the dilaton condense a mass-term for the (axial-)vector d.o.f. arises:
\begin{equation}
m_{1}^{2}=bG_{0}^{2}\text{ .} \label{m1}%
\end{equation}
$\mathcal{L}_{3,AV}$ and $\mathcal{L}_{4,AV}$ are:%
\begin{equation}
\mathcal{L}_{3,AV}=-2ig_{2}\left(  \mathrm{Tr}[L_{\mu\nu}[L^{\mu},L^{\nu
}]]+\mathrm{Tr}[R_{\mu\nu}[R^{\mu},R^{\nu}]]\right)  \text{ ,}%
\end{equation}%
\begin{align}
\mathcal{L}_{4,AV}  &  =g_{3}\left\{  \mathrm{Tr}\left[  L^{\mu}L^{\nu}L_{\mu
}L_{\nu}\right]  +\mathrm{Tr}\left[  R^{\mu}R^{\nu}R_{\mu}R_{\nu}\right]
\right\}  +\nonumber\\
&  g_{4}\left\{  \mathrm{Tr}\left[  L^{\mu}L_{\mu}L^{\nu}L_{\nu}\right]
+\mathrm{Tr}\left[  R^{\mu}R_{\mu}R^{\nu}R_{\nu}\right]  \right\} \nonumber\\
&  g_{5}\mathrm{Tr}\left[  R^{\mu}R_{\mu}\right]  \mathrm{Tr}\left[  L^{\mu
}L_{\mu}\right]  +\nonumber\\
&  g_{6}\left\{  \mathrm{Tr}[L_{\mu}L^{\mu}]\mathrm{Tr}[L_{\mu}L^{\mu
}]+\mathrm{Tr}[R_{\mu}R^{\mu}]\mathrm{Tr}[R_{\mu}R^{\mu}]\right\}  \text{ .}%
\end{align}
$\mathcal{L}_{4,AV}$ does not influence decays and is not relevant here. Finally:%

\begin{align}
\mathcal{L}_{\Phi,AV}  &  =\mathrm{Tr}\left[  (ig_{1}(\Phi R^{\mu}-L^{\mu}%
\Phi))^{\dagger}(\partial^{\mu}\Phi)\right]  +\mathrm{Tr}\left[
(\partial^{\mu}\Phi)^{\dagger}(ig_{1}(\Phi R^{\mu}-L^{\mu}\Phi))\right]
\nonumber\\
&  +\mathrm{Tr}\left[  (ig_{1}(\Phi R^{\mu}-L^{\mu}\Phi))^{\dagger}%
(ig_{1}(\Phi R^{\mu}-L^{\mu}\Phi))\right] \nonumber\\
&  +\frac{h_{1}}{2}\mathrm{Tr}\left[  \Phi\Phi^{\dag}\right]  Tr\left[
L_{\mu}L^{\mu}+R_{\mu}R^{\mu}\right] \nonumber\\
&  +h_{2}\mathrm{Tr}\left[  \Phi^{\dag}L_{\mu}L^{\mu}\Phi+\Phi R_{\mu}R^{\mu
}\Phi\right]  +2h_{3}\mathrm{Tr}\left[  \Phi R_{\mu}\Phi^{\dag}L^{\mu}\right]
\text{ }.
\end{align}
The large-$N_{c}$ dependence of the parameters is:%
\begin{align*}
g_{1},\text{ }g_{2},\text{ }g  &  \propto N_{c}^{-1/2}\;,\\
\text{ }h_{2},\text{ }h_{3},\text{ }g_{3},\text{ }g_{4}  &  \propto N_{c}%
^{-1}\;,\\
\text{ }h_{1},\text{ }g_{5},\text{ }g_{6}  &  \propto N_{c}^{-2}\;,
\end{align*}%
\begin{equation}
b\propto N_{c}^{-2}\rightarrow m_{1}^{2}\propto N_{c}^{0}\;.
\end{equation}

\subsubsection{Explicit breaking of chiral symmetry through bare quark masses}

The nonzero quark masses is taken into account by the term:%
\begin{equation}
\mathrm{Tr}[H(\Phi^{\dag}+\Phi)]\text{ ,}%
\end{equation}%
\begin{equation}
H=diag\{h_{0}^{1},h_{0}^{2},...,h_{0}^{N_{f}}\} \label{h0i}%
\end{equation}
with $h_{0}^{k}\propto m_{k}$ and $h_{0}^{k}\propto N_{c}^{1/2}$. The effect
of this object was already studied in Sec. 3.2.3. A unique minimum of the
potential exists.

Further chiral breaking terms are given by%
\begin{align}
&  \mathrm{Tr}[\varepsilon\Phi^{\dag}\Phi]\text{ ,}\label{vecexpl}\\
&  \frac{1}{2}\mathrm{Tr}[\delta(L^{\mu})^{2}+(R^{\mu})^{2}]\text{ ,}%
\end{align}
where $\varepsilon$ and $\delta$ are diagonal matrices whose $k$-element is
proportional to $m_{k}^{2}.$ These terms generate a standard contribution to
the meson masses. Indeed, when $\varepsilon$ and $\delta$ are proportional to
the identity, these terms can be absorbed away and have no influence on the
results. It is then the difference of masses which is important here.

Finally the explicit symmetry breaking Lagrangian is given by%
\begin{equation}
\mathcal{L}_{eSB}=\mathrm{Tr}[H(\Phi^{\dag}+\Phi)]+\mathrm{Tr}[\varepsilon
\Phi^{\dag}\Phi]+\frac{1}{2}\mathrm{Tr}[\delta(L^{\mu})^{2}+(R^{\mu}%
)^{2}]\text{ .}%
\end{equation}

\subsubsection{The whole Lagrangian of the eLSM}

We now put all the elements together and finally obtain the Lagrangian of the
extended Linear Sigma Model in the mesonic sector:%

\begin{equation}
\mathcal{L}_{eLSM}=\mathcal{L}_{G\Phi}+\mathcal{L}_{AV}+\mathcal{L}_{U_{A}%
(1)}+\mathcal{L}_{eSB}\text{ .}%
\end{equation}
Explicitly:
\begin{align}
\mathcal{L}_{eLSM} &  =\frac{1}{2}(\partial_{\mu}G)^{2}-\frac{1}{4}\frac
{m_{G}^{2}}{\Lambda_{G}^{2}}\left[  G^{4}\ln\left(  \frac{G}{\Lambda_{G}%
}\right)  -\frac{G^{4}}{4}\right]  \nonumber\\
&  +\mathrm{Tr}\left[  (D^{\mu}\Phi)^{\dagger}(D_{\mu}\Phi)-aG^{2}\Phi^{\dag
}\Phi-\lambda_{2}\left(  \Phi^{\dag}\Phi\right)  ^{2}\right]  -\lambda
_{1}(\mathrm{Tr}[\Phi^{\dag}\Phi])^{2}\nonumber\\
&  +c_{1}(\det\Phi^{\dag}-\det\Phi)^{2}\text{ }+\mathrm{Tr}[H(\Phi^{\dag}%
+\Phi)]+\mathrm{Tr}[\varepsilon\Phi^{\dag}\Phi]+\text{.}\nonumber\\
&  -\frac{1}{4}\mathrm{Tr}[(L^{\mu\nu})^{2}+(R^{\mu\nu})^{2}]+\frac{b}{2}%
G^{2}\mathrm{Tr}[(L^{\mu})^{2}+(R^{\mu})^{2}]\nonumber\\
&  +\frac{1}{2}\mathrm{Tr}[\delta(L^{\mu})^{2}+(R^{\mu})^{2}]\text{
}\nonumber\\
&  -2ig_{2}\left(  \mathrm{Tr}[L_{\mu\nu}[L^{\mu},L^{\nu}]]+\mathrm{Tr}%
[R_{\mu\nu}[R^{\mu},R^{\nu}]]\right)  \nonumber\\
&  +\frac{h_{1}}{2}\mathrm{Tr}\left[  \Phi\Phi^{\dag}\right]  Tr\left[
L_{\mu}L^{\mu}+R_{\mu}R^{\mu}\right]  +\nonumber\\
&  +h_{2}\mathrm{Tr}\left[  \Phi^{\dag}L_{\mu}L^{\mu}\Phi+\Phi R_{\mu}R^{\mu
}\Phi\right]  +2h_{3}\mathrm{Tr}\left[  \Phi R_{\mu}\Phi^{\dag}L^{\mu}\right]
\text{ }+...\text{,}\label{ltot}%
\end{align}
with
\begin{equation}
D^{\mu}\Phi=\partial^{\mu}\Phi-ig_{1}(L^{\mu}\Phi-\Phi R^{\mu})\text{ .}%
\end{equation}
This is a realistic model of low-energy QCD which includes from the very
beginning (pseudo)scalar and (axial-)vector quarkonia and one scalar glueball.
Next, we show the comparison with data.

\subsection{\emph{Results for the case }$N_{f}=3$}

In Ref. \cite{dick} the case $N_{f}=3$ has been studied in detail. In the
first step, the glueball $G$ is frozen: we simply set $G=G_{0}$ and do not
consider its fluctuations (this is done later on). In Table 4 we show the
assignments of the fields entering in model and their PDG counterparts.

\begin{center}
\textbf{Table 4}: Fields and correspondence to PDG%

\begin{tabular}
[c]{|c|c|c|c|c|c|}\hline
Field & PDG & Quark content & $I$ & $J^{PC}$ & Mass (MeV)\\\hline
$\pi^{+},\pi^{-},\pi^{0}$ & $\pi$ & $u\bar{d},d\bar{u},\frac{u\bar{u}-d\bar
{d}}{\sqrt{2}}$ & $1$ & $0^{-+}$ & $139.57$\\\hline
$K^{+},K^{-},K^{0},\bar{K}^{0}$ & $K$ & $u\bar{s},s\bar{u},d\bar{s},s\bar{d}$
& $1/2$ & $0^{-+}$ & $493.677$\\\hline
$\eta_{N}a+\eta_{S}b$ & $\eta$ & $\frac{u\bar{u}+d\bar{d}}{\sqrt{2}}a+s\bar
{s}b$ & $0$ & $0^{-+}$ & $547.86$\\\hline
$-\eta_{N}a+\eta_{S}b$ & $\eta^{\prime}(958)$ & $\frac{u\bar{u}+d\bar{d}%
}{\sqrt{2}}a+s\bar{s}b$ & $0$ & $0^{-+}$ & $957.78$\\\hline
$a_{0}^{+},a_{0}^{-},a_{0}^{0}$ & $a_{0}(1450)$ & $u\bar{d},d\bar{u}%
,\frac{u\bar{u}-d\bar{d}}{\sqrt{2}}$ & $1$ & $0^{++}$ & $1474$\\\hline
$K_{S}^{+},K_{S}^{-},K_{S}^{0},\bar{K}_{S}^{0}$ & $K_{0}^{\ast}(1430)$ &
$u\bar{s},s\bar{u},d\bar{s},s\bar{d}$ & $1/2$ & $0^{++}$ & $1425$\\\hline
$\sigma_{N}$ & $f_{0}(1370)$ & $\frac{u\bar{u}+d\bar{d}}{\sqrt{2}}$ & $0$ &
$0^{++}$ & $1350$\\\hline
$\sigma_{S}$ & $f_{0}(1500)$ & $s\bar{s}$ & $0$ & $0^{++}$ & $1504$\\\hline
$\rho^{+},\rho^{-},\rho^{0}$ & $\rho(770)$ & $u\bar{d},d\bar{u},\frac{u\bar
{u}-d\bar{d}}{\sqrt{2}}$ & $1$ & $1^{--}$ & $775.26$\\\hline
$K^{\ast+},K^{\ast-},K^{\ast0},\bar{K}^{\ast0}$ & $K^{\ast}(892)$ & $u\bar
{s},s\bar{u},d\bar{s},s\bar{d}$ & $1/2$ & $1^{--}$ & $891.86$\\\hline
$\omega_{N}$ & $\omega(782)$ & $\frac{u\bar{u}+d\bar{d}}{\sqrt{2}}$ & $0$ &
$1^{--}$ & $782.65$\\\hline
$\omega_{S}$ & $\phi(1020)$ & $s\bar{s}$ & $0$ & $1^{--}$ & $1019.461$\\\hline
$a_{1}^{+},a_{1}^{-},a_{1}^{0}$ & $a_{1}(1230)$ & $u\bar{d},d\bar{u}%
,\frac{u\bar{u}-d\bar{d}}{\sqrt{2}}$ & $1$ & $1^{++}$ & $1230$\\\hline
$K_{1}^{+},K_{1}^{-},K_{1}^{0},\bar{K}_{1}^{0}$ & $K_{1}(1270)$ & $u\bar
{s},s\bar{u},d\bar{s},s\bar{d}$ & $1/2$ & $1^{++}$ & $1272$\\\hline
$f_{1,N}$ & $f_{1}(1285)$ & $\frac{u\bar{u}+d\bar{d}}{\sqrt{2}}$ & $0$ &
$1^{++}$ & $1281.9$\\\hline
$f_{1,S}$ & $f_{1}(1420)$ & $s\bar{s}$ & $0$ & $1^{++}$ & $1426.4$\\\hline
\end{tabular}

\bigskip

\bigskip
\end{center}

The masses of the particles were calculated in Refs. \cite{dick,denisdiss}.
The procedure is the same of Sec. 3.2.3, just a bit longer because many fields
are present.

The pseudoscalar masses are:%

\begin{align}
m_{\pi}^{2}  &  =Z_{\pi}^{2}\left[  m_{0}^{2}+\left(  \lambda_{1}%
+\frac{\lambda_{2}}{2}\right)  \phi_{N}^{2}+\lambda_{1}\phi_{S}^{2}\right]
\equiv\frac{Z_{\pi}^{2}h_{0N}}{\phi_{N}}\text{ ,}\\
m_{K}^{2}  &  =Z_{K}^{2}\left[  m_{0}^{2}+\left(  \lambda_{1}+\frac
{\lambda_{2}}{2}\right)  \phi_{N}^{2}-\frac{\lambda_{2}}{\sqrt{2}}\phi_{N}%
\phi_{S}+\left(  \lambda_{1}+\lambda_{2}\right)  \phi_{S}^{2}\right]  \text{
,}\\
m_{\eta_{N}}^{2}  &  =Z_{\pi}^{2}\left[  m_{0}^{2}+\left(  \lambda_{1}%
+\frac{\lambda_{2}}{2}\right)  \phi_{N}^{2}+\lambda_{1}\phi_{S}^{2}%
+c_{1}\,\phi_{N}^{2}\phi_{S}^{2}\right] \nonumber\\
&  \equiv Z_{\pi}^{2}\left(  \frac{h_{0N}}{\phi_{N}}+c_{1}\,\phi_{N}^{2}%
\phi_{S}^{2}\right)  \;,\\
m_{\eta_{S}}^{2}  &  =Z_{\eta_{S}}^{2}\left[  m_{0}^{2}+\lambda_{1}\phi
_{N}^{2}+\left(  \lambda_{1}+\lambda_{2}\right)  \phi_{S}^{2}+\frac{c_{1}}%
{4}\phi_{N}^{4}\right] \nonumber\\
&  \equiv Z_{\eta_{S}}^{2}\left(  \frac{h_{0S}}{\phi_{S}}+\frac{c_{1}}{4}%
\phi_{N}^{4}\right)  \text{ ,}\\
m_{\eta_{NS}}^{2}  &  =Z_{\pi}Z_{\pi_{S}}\frac{c_{1}}{2}\phi_{N}^{3}\phi
_{S}\text{ ,}%
\end{align}
where $m_{\eta_{NS}}^{2}$ is a mixing term, leading to:%
\begin{equation}
\;m_{\eta^{\prime}/\eta}^{2}=\frac{1}{2}\left[  m_{\eta_{N}}^{2}+m_{\eta_{S}%
}^{2}\pm\sqrt{(m_{\eta_{N}}^{2}-m_{\eta_{S}}^{2})^{2}+4m_{\eta_{NS}}^{4}%
}\right]  \;. \label{eq:eta}%
\end{equation}
The constants $Z_{k}$ arise when unphysical axial-vector-pseudoscalar mixing
terms are eliminated by proper shifts and renormalization constants, see
details in Ref. \cite{dick}. They read:%
\begin{align}
Z_{\pi}  &  =Z_{\eta_{N}}=\frac{m_{a_{1}}}{\sqrt{m_{a_{1}}^{2}-g_{1}^{2}%
\phi_{N}^{2}}}\text{ ,} & Z_{K}  &  =\frac{2m_{K_{1}}}{\sqrt{4m_{K_{1}}%
^{2}-g_{1}^{2}(\phi_{N}+\sqrt{2}\phi_{S})^{2}}}\text{ ,}\label{Z_pi}\\
Z_{\eta_{S}}  &  =\frac{m_{f_{1S}}}{\sqrt{m_{f_{1S}}^{2}-2g_{1}^{2}\phi
_{S}^{2}}}\text{ ,} & Z_{K_{0}^{\star}}  &  =\frac{2m_{K^{\star}}}%
{\sqrt{4m_{K^{\star}}^{2}-g_{1}^{2}(\phi_{N}-\sqrt{2}\phi_{S})^{2}}}\;.
\label{Z_K_S}%
\end{align}

The masses of the scalar mesons are:
\begin{align}
m_{a_{0}}^{2}  &  =m_{0}^{2}+\left(  \lambda_{1}+\frac{3}{2}\lambda
_{2}\right)  \phi_{N}^{2}+\lambda_{1}\phi_{S}^{2}\text{ ,}\label{m_a_0}\\
m_{K_{0}^{\star}}^{2}  &  =Z_{K_{0}^{\star}}^{2}\left[  m_{0}^{2}+\left(
\lambda_{1}+\frac{\lambda_{2}}{2}\right)  \phi_{N}^{2}+\frac{\lambda_{2}%
}{\sqrt{2}}\phi_{N}\phi_{S}+\left(  \lambda_{1}+\lambda_{2}\right)  \phi
_{S}^{2}\right]  \text{ ,}\\
m_{\sigma_{N}}^{2}  &  =m_{0}^{2}+3\left(  \lambda_{1}+\frac{\lambda_{2}}%
{2}\right)  \phi_{N}^{2}+\lambda_{1}\phi_{S}^{2}\text{ ,}\label{eq:sigN}\\
m_{\sigma_{S}}^{2}  &  =m_{0}^{2}+\lambda_{1}\phi_{N}^{2}+3\left(  \lambda
_{1}+\lambda_{2}\right)  \phi_{S}^{2}\text{ .} \label{eq:sigNS}%
\end{align}

The masses of the vector mesons are:%

\begin{align}
m_{\rho}^{2}  &  =m_{1}^{2}+\frac{1}{2}(h_{1}+h_{2}+h_{3})\phi_{N}^{2}%
+\frac{h_{1}}{2}\phi_{S}^{2}+2\delta_{N}\;,\label{m_rho}\\
m_{K^{\star}}^{2}  &  =m_{1}^{2}+\frac{1}{4}\left(  g_{1}^{2}+2h_{1}%
+h_{2}\right)  \phi_{N}^{2}\nonumber\\
&  +\frac{1}{\sqrt{2}}\phi_{N}\phi_{S}(h_{3}-g_{1}^{2})+\frac{1}{2}(g_{1}%
^{2}+h_{1}+h_{2})\phi_{S}^{2}+\delta_{N}+\delta_{S}\;,\\
m_{\omega_{N}}^{2}  &  =m_{\rho}^{2}\;,\\
m_{\omega_{S}}^{2}  &  =m_{1}^{2}+\frac{h_{1}}{2}\phi_{N}^{2}+\left(
\frac{h_{1}}{2}+h_{2}+h_{3}\right)  \phi_{S}^{2}+2\delta_{S}\;,
\end{align}

The masses of axial-vector mesons are:%

\begin{align}
m_{a_{1}}^{2}  &  =m_{1}^{2}+\frac{1}{2}(2g_{1}^{2}+h_{1}+h_{2}-h_{3})\phi
_{N}^{2}+\frac{h_{1}}{2}\phi_{S}^{2}+2\delta_{N}\;,\label{m_a_1}\\
m_{K_{1}}^{2}  &  =m_{1}^{2}+\frac{1}{4}\left(  g_{1}^{2}+2h_{1}+h_{2}\right)
\phi_{N}^{2}\nonumber\\
&  -\frac{1}{\sqrt{2}}\phi_{N}\phi_{S}(h_{3}-g_{1}^{2})+\frac{1}{2}\left(
g_{1}^{2}+h_{1}+h_{2}\right)  \phi_{S}^{2}+\delta_{N}+\delta_{S}\;,\\
m_{f_{1N}}^{2}  &  =m_{a_{1}}^{2}\;,\\
m_{f_{1S}}^{2}  &  =m_{1}^{2}+\frac{h_{1}}{2}\phi_{N}^{2}+\left(  2g_{1}%
^{2}+\frac{h_{1}}{2}+h_{2}-h_{3}\right)  \phi_{S}^{2}+2\delta_{S}\;.
\label{m_f1_S}%
\end{align}

The potential depends on two condensates $\phi_{N}\equiv\left\langle \bar
{u}u+\bar{d}d\right\rangle /\sqrt{2},$ $\phi_{S}\equiv\left\langle
ss\right\rangle $, the light-quark condensate and the strange quark
condensate, respectively (the discussion is still very similar to Sec. 3.2.3):%

\begin{align}
\mathcal{V}(\phi_{N},\phi_{S})  &  =\frac{1}{2}m_{0}^{2}(\phi_{N}^{2}+\phi
_{S}^{2})+\frac{\lambda_{1}}{4}(\phi_{N}^{4}+2\phi_{N}^{2}\phi_{S}^{2}%
+\phi_{S}^{4})\nonumber\\
&  +\frac{\lambda_{2}}{4}\left(  \frac{\phi_{N}^{4}}{2}+\phi_{S}^{4}\right)
-h_{0N}\phi_{N}-h_{0S}\phi_{S}\text{ .} \label{eq:Pot}%
\end{align}
One obtains$\ \phi_{N}$ and $\phi_{S}\ $upon minimization$:$
\begin{align}
\frac{\partial\mathcal{V}(\phi_{N},\phi_{S})}{\partial\phi_{N}}\overset{!}%
{=}0  &  \Leftrightarrow h_{0N}=[m_{0}^{2}+\lambda_{1}(\phi_{N}^{2}+\phi
_{S}^{2})]\phi_{N}+\frac{\lambda_{2}}{2}\phi_{N}^{3}\text{ ,}\\
\frac{\partial\mathcal{V}(\phi_{N},\phi_{S})}{\partial\phi_{S}}\overset{!}%
{=}0  &  \Leftrightarrow h_{0S}=[m_{0}^{2}+\lambda_{1}(\phi_{N}^{2}+\phi
_{S}^{2})]\phi_{S}+\lambda_{2}\phi_{S}^{3}\text{ .}%
\end{align}

In summary, the masses depend on the two chiral condensates. It is easy to see
that particles come in chiral partners, which become degenerate when the
condensates vanish. The situation is indeed in principle very similar to Sec.
3.2.3, although much more complicated from a technical point of view. The
calculation of decays can be carried out in a straightforward way, which
however requires some hard work \cite{dick,denisdiss}.

\bigskip

\begin{center}
\textbf{Table 5 :} Values of the parameters (from \cite{dick}).%

\begin{equation}%
\begin{tabular}
[c]{|l|l|}\hline
Parameter & Value\\\hline
$m_{0}^{2}$ $\left[  \text{GeV}^{2}\right]  $ & $-0.9183\pm0.0006$\\\hline
$m_{1}^{2}$ $\left[  \text{GeV}^{2}\right]  $ & $0.4135\pm0.0147$\\\hline
$c_{1}$ $\left[  \text{GeV}^{-2}\right]  $ & $450.5420\pm7.033$\\\hline
$\delta_{S}$ $\left[  \text{GeV}^{2}\right]  $ & $0.1511\pm0.0038$\\\hline
$g_{1}$ & $5.843\pm0.018$\\\hline
$g_{2}$ & $3.0250\pm0.2329$\\\hline
$\phi_{N}$ $\left[  \text{GeV}\right]  $ & $0.1646\pm0.0001$\\\hline
$\phi_{S}$ $\left[  \text{GeV}\right]  $ & $0.1262\pm0.0001$\\\hline
$h_{2}$ & $9.8796\pm0.6627$\\\hline
$h_{3}$ & $4.8667\pm0.0864$\\\hline
$\lambda_{2}$ & $68.2972\pm0.0435$\\\hline
\end{tabular}
\ \ \ \ \ \ \label{Tab:Parameter}%
\end{equation}

\end{center}

\bigskip\newpage

\begin{center}
\textbf{Table 6:} Results of the fit (from \cite{dick})
\end{center}

\begin{equation}%
\begin{tabular}
[c]{|c|c|c|}\hline
Observable & Fit $\left[  \text{MeV}\right]  $ & Experiment $\left[
\text{MeV}\right]  $\\\hline
$f_{\pi}$ & $96.3\pm0.7$ & $92.2\pm4.6$\\\hline
$f_{K}$ & $106.9\pm0.6$ & $110.4\pm5.5$\\\hline
$m_{\pi}$ & $141.0\pm5.8$ & $137.3\pm6.9$\\\hline
$m_{K}$ & $485.6\pm3.0$ & $495.6\pm24.8$\\\hline
$m_{\eta}$ & $509.4\pm3.0$ & $547.9\pm27.4$\\\hline
$m_{\eta^{\prime}}$ & $962.5\pm5.6$ & $957.8\pm47.9$\\\hline
$m_{\rho}$ & $783.1\pm7.0$ & $775.5\pm38.8$\\\hline
$m_{K^{\star}}$ & $885.1\pm6.3$ & $893.8\pm44.7$\\\hline
$m_{\phi}$ & $975.1\pm6.4$ & $1019.5\pm51.0$\\\hline
$m_{a_{1}}$ & $1186\pm6.0$ & $1230\pm62$\\\hline
$m_{f_{1}\left(  1420\right)  }$ & $1372.4\pm5.3$ & $1426\pm71$\\\hline
$m_{a_{0}}$ & $1363\pm1$ & $1474\pm74$\\\hline
$m_{K_{0}^{\star}}$ & $1450\pm1$ & $1425\pm71$\\\hline
$\Gamma_{\rho\rightarrow\pi\pi}$ & $160.9\pm4.4$ & $149.1\pm7.4$\\\hline
$\Gamma_{K^{\star}\rightarrow K\pi}$ & $44.6\pm1.9$ & $46.2\pm2.3$\\\hline
$\Gamma_{\phi\rightarrow\bar{K}K}$ & $3.34\pm0.14$ & $3.54\pm0.18$\\\hline
$\Gamma_{a_{1}\rightarrow\rho\pi}$ & $549\pm43$ & $425\pm175$\\\hline
$\Gamma_{a_{1}\rightarrow\pi\gamma}$ & $0.66\pm0.01$ & $0.64\pm0.25$\\\hline
$\Gamma_{f_{1}\left(  1420\right)  \rightarrow K^{\star}K}$ & $44.6\pm39.9$ &
$43.9\pm2.2$\\\hline
$\Gamma_{a_{0}}$ & $266\pm12$ & $265\pm13$\\\hline
$\Gamma_{K_{0}^{\star}\rightarrow K\pi}$ & $285\pm12$ & $270\pm80$\\\hline
\end{tabular}
\ \ \ \ \nonumber\label{Tab:Obser}%
\end{equation}

Eleven parameters, listed in Table 5, were fitted to 21 experimental
quantities listed in Table 6. (Note: when the experimental error was smaller
than 5\% of the measured value, the error considered in the fit was set to
5\%. This is because our model, which neglects in the present form isospin
breaking, cannot achieve a higher precision).

Following considerations hold:

(i) The scalar quark-antiquark mesons lie above 1 GeV (see Table 6).

(ii) As a consequence of (i), the light scalar mesons with a mass below 1 GeV
are something else (four-quark states, see the next section).

(iii) In Ref. \cite{dick} various consequences of the fit were investigated.
For instance, the branching ratios of $a_{0}$ are in good agreement with the
experimental data.

(iv) The spectral function of the $\rho$ and the $a_{1}$ meson can be
successfully calculated \cite{anja}.

(v) The eLSM has also been applied to baryons in the so-called mirror
assignment \cite{susanna}.

(vi) Studies of the model at nonzero density have been performed in\ Refs.
\cite{susagiu,achimlast}.

(vii) Some preliminary attempts to study the eLSM at nonzero temperature have
been undertaken in Refs. \cite{achim,achimnc}.

(viii) The model has been successfully extended to the case $N_{f}=4$
\cite{walaa}, in which charmed mesons are included.

\section{Glueballs in the eLSM}

According to lattice QCD many glueballs with various quantum numbers should
exist, see Ref. \cite{mainlattice,morningstar,vento} and Table 7 (for the
uncertainties, see Ref. \cite{mainlattice}). However, up to now no glueball
state has been unambiguously identified (although for some of them some
candidates exist). The eSLM introduced in the previous section can help to
elucidate some features of glueballs. In fact, the scalar glueball is a very
important building block of the model and other glueballs, such as the
pseudoscalar one, can be easily coupled to the eLSM.

\begin{center}
\textbf{Table 7}: Central values of glueball masses from lattice (from
\cite{mainlattice}).%

\begin{tabular}
[c]{|c|c|}\hline
$J^{PC}$ & Value [GeV]\\\hline
$0^{++}$ & $\,1.70$\\\hline
$2^{++}$ & $2.39$\\\hline
$0^{-+}$ & $2.55$\\\hline
$1^{-+}$ & $2.96$\\\hline
$2^{-+}$ & $3.04$\\\hline
$3^{+-}$ & $3.60$\\\hline
$3^{++}$ & $3.66$\\\hline
$1^{--}$ & $3.81$\\\hline
$2^{--}$ & $4.0$\\\hline
$3^{--}$ & $4.19$\\\hline
$2^{+-}$ & $4.22$\\\hline
$0^{+-}$ & $4.77$\\\hline
\end{tabular}

\end{center}

\subsection{Scalar glueball}

The scalar glueball is the lightest gluonic state predicted by lattice\ QCD
and is a natural element of the eLSM as the excitation of the dilaton
field\ \cite{dick,stanilast}. The eLSM makes predictions for the lightest
glueball state in a chiral framework, completing previous phenomenological
works on the subject \cite{close,longglueball,cheng,gutsche}. The result of
the recent study of Ref. \cite{stanilast} shows that the scalar glueball is
predominately contained in the resonance $f_{0}(1710)$ This assignment is in
agreement with the old lattice result of Ref. \cite{weingarten}, with the
recent lattice result of Ref. \cite{chenlattice}, with other hadronic
approaches \cite{longglueball,cheng} and lately also within an holographic
approach \cite{rehban}.

The admixtures as calculated in Ref. \cite{stanilast} read:%

\begin{equation}
\left(
\begin{array}
[c]{c}%
f_{0}(1370)\\
f_{0}(1500)\\
f_{0}(1710)
\end{array}
\right)  =\left(
\begin{array}
[c]{ccc}%
-0.91 & 0.24 & -0.33\\
0.30 & 0.94 & -0.17\\
-0.27 & 0.26 & 0.93
\end{array}
\right)  \left(
\begin{array}
[c]{c}%
\sigma_{N}\\
\sigma_{S}\\
G
\end{array}
\right)  \text{ }. \label{mixmat}%
\end{equation}
The following admixtures of the bare fields to the resonances follow:
\begin{align}
f_{0}(1370):  &  \quad83\%\,\sigma_{N}\;,\quad6\%\,\sigma_{S}\;,\quad
11\%\,G\text{ },\nonumber\\
f_{0}(1500):  &  \quad9\%\,\sigma_{N}\;,\quad88\%\,\sigma_{S}\;,\quad
3\%\,G\text{ },\label{admix1}\\
f_{0}(1710):  &  \quad8\%\,\sigma_{N}\;,\quad6\%\,\sigma_{S}\;,\quad
86\%\,G\text{ }.\nonumber
\end{align}
The additional 5 parameters listed in Table\ 8 are needed in the scalar
isoscalar sector. They were fitted to the experimental quantities of Table 9.
Further predictions of the approach are reported in Table 10.

\begin{center}
\textbf{Table 8}: Additional parameters for the $J^{PC}=0^{++}$ sector (from
\cite{stanilast}).%

\begin{tabular}
[c]{|c|c|}\hline
Parameter & Value\\\hline
$\Lambda$ & $3297\,[$MeV$]$\\\hline
$m_{G}$ & $1525\,[$MeV$]$\\\hline
$\lambda_{1}$ & $6.25$\\\hline
$h_{1}$ & $-3.22$\\\hline
$\epsilon_{S}$ & $0.4212\times10^{6}\,[$MeV$^{2}]$\\\hline
\end{tabular}

\end{center}

\bigskip

\begin{center}
\textbf{Table 9}: Fitted quantities in the $J^{PC}=0^{++}$ sector (from
\cite{stanilast}).%

\begin{tabular}
[c]{|c|c|c|}\hline
Quantity & Fit [MeV] & Exp. [MeV]\\\hline
$M_{f_{0}(1370)}$ & $1444$ & $1200$-$1500$\\\hline
$M_{f_{0}(1500)}$ & $1534$ & $1505\pm6$\\\hline
$M_{f_{0}(1710)}$ & $1750$ & $1720\pm6$\\\hline
$f_{0}(1370)\rightarrow\pi\pi$ & $423.6$ & -\\\hline
$f_{0}(1500)\rightarrow\pi\pi$ & $39.2$ & $38.04\pm4.95$\\\hline
$f_{0}(1500)\rightarrow K\bar{K}$ & $9.1$ & $9.37\pm1.69$\\\hline
$f_{0}(1710)\rightarrow\pi\pi$ & $28.3$ & $29.3\pm6.5$\\\hline
$f_{0}(1710)\rightarrow K\bar{K}$ & $73.4$ & $71.4\pm29.1$\\\hline
\end{tabular}

\end{center}

\bigskip

In our solution, $f_{0}(1370)$ decays predominantly into two pions with a
decay width of about $400$ MeV, in agreement with its interpretation as
predominantly nonstrange $\bar{q}q$ state\ \cite{dick,stanilast}. This is in
qualitative agreement with the experimental analysis of Ref.\ \cite{bugg},
where $\Gamma_{f_{0}(1370)\rightarrow\pi\pi}=325$ MeV, $\Gamma_{f_{0}%
(1370)\rightarrow4\pi}\approx50$ MeV, and $\Gamma_{f_{0}(1370)\rightarrow
\eta\eta}/\Gamma_{f_{0}(1370)\rightarrow\pi\pi}=0.19\pm0.07$. Moreover, the
decay channel $f_{0}(1500)\rightarrow\eta\eta$ is in good agreement with PDG,
while the decay channel $f_{0}(1710)\rightarrow\eta\eta$ is slightly larger
than the experiment.

\bigskip

\begin{center}
\textbf{Table 10:} further properties in the $J^{PC}=0^{++}$ sector (from
\cite{stanilast}).

$%
\begin{tabular}
[c]{|c|c|c|}\hline
Decay Channel & Our Value [MeV] & Exp. [MeV]\\\hline
$f_{0}(1370)\rightarrow K\bar{K}$ & $117.5$ & -\\\hline
$f_{0}(1370)\rightarrow\eta\eta$ & $43.3$ & -\\\hline
$f_{0}(1370)\rightarrow\rho\rho\rightarrow4\pi$ & $13.8$ & -\\\hline
$f_{0}(1500)\rightarrow\eta\eta$ & $4.7$ & $5.56\pm1.34$\\\hline
$f_{0}(1500)\rightarrow\rho\rho\rightarrow4\pi$ & $0.2$ & $>54.0\pm
7.1$\\\hline
$f_{0}(1710)\rightarrow\eta\eta$ & $57.9$ & $34.3\pm17.6$\\\hline
$f_{0}(1710)\rightarrow\rho\rho\rightarrow4\pi$ & $0.5$ & -\\\hline
\end{tabular}
\ \ \ \ $
\end{center}

\bigskip

In conclusion, present results show that $f_{0}(1710)$ is a very good scalar
glueball candidate.

\bigskip

\subsection{The pseudoscalar glueball}

The pseudoscalar glueball is related to the chiral anomaly and couples in a
chirally invariant way to light mesons \cite{psg} and also to baryons
\cite{psgproc}. The chiral Lagrangian which couples the pseudoscalar glueball
$\tilde{G}\equiv\left\vert gg\right\rangle $ with quantum numbers
$J^{PC}=0^{-+}$ to scalar and pseudoscalar mesons reads explicitly:
\begin{equation}
\mathcal{L}_{\tilde{G}}^{int}=ic_{\tilde{G}\Phi}\tilde{G}\left(
\text{\textrm{det}}\Phi-\text{\textrm{det}}\Phi^{\dag}\right)  \text{ .}
\label{intlag}%
\end{equation}
The branching ratios of $\tilde{G}$ are reported in Table 11 for two choices
of the pseudoscalar masses: $M_{\tilde{G}}=2.6$ GeV predicted by lattice QCD
\cite{mainlattice} and $M_{\tilde{G}}=2.37$ GeV, in agreement with the
pseudoscalar particle $X(2370)$ measured by BES \cite{bespsg} (this resonance
could be the pseudoscalar glueball). The branching ratios are presented
relative to the total decay width of the pseudoscalar glueball $\Gamma
_{\tilde{G}}^{tot}$.

\bigskip

\newpage

\begin{center}
\textbf{Table 11}: Branching ratios for the pseudoscalar glueball (from
\cite{psg}).%

\begin{tabular}
[c]{|c|c|c|}\hline
Quantity & Case (i): $M_{\tilde{G}}=2.6$ GeV & Case (ii): $M_{\tilde{G}}=2.37$
GeV\\\hline
$\Gamma_{\tilde{G}\rightarrow KK\eta}/\Gamma_{\tilde{G}}^{tot}$ & $0.049$ &
$0.043$\\\hline
$\Gamma_{\tilde{G}\rightarrow KK\eta^{\prime}}/\Gamma_{\tilde{G}}^{tot}$ &
$0.019$ & $0.011$\\\hline
$\Gamma_{\tilde{G}\rightarrow\eta\eta\eta}/\Gamma_{\tilde{G}}^{tot}$ & $0.016$
& $0.013$\\\hline
$\Gamma_{\tilde{G}\rightarrow\eta\eta\eta^{\prime}}/\Gamma_{\tilde{G}}^{tot}$
& $0.0017$ & $0.00082$\\\hline
$\Gamma_{\tilde{G}\rightarrow\eta\eta^{\prime}\eta^{\prime}}/\Gamma_{\tilde
{G}}^{tot}$ & $0.00013$ & $0$\\\hline
$\Gamma_{\tilde{G}\rightarrow KK\pi}/\Gamma_{\tilde{G}}^{tot}$ & $0.47$ &
$0.47$\\\hline
$\Gamma_{\tilde{G}\rightarrow\eta\pi\pi}/\Gamma_{\tilde{G}}^{tot}$ & $0.16$ &
$0.17$\\\hline
$\Gamma_{\tilde{G}\rightarrow\eta^{\prime}\pi\pi}/\Gamma_{\tilde{G}}^{tot}$ &
$0.095$ & $0.090$\\\hline
\end{tabular}

\end{center}

In conclusion, we predict that $KK\pi$ is the dominant decay channel (50\%),
followed by sizable $\eta\pi\pi$ and $\eta^{\prime}\pi\pi$ decay channels
(16\% and 10\% respectively). Moreover, we also predict the decay into three
pions should vanish:%
\begin{equation}
\Gamma_{\tilde{G}\rightarrow\pi\pi\pi}=0\text{ .}%
\end{equation}
These are simple and testable theoretical predictions which can be helpful in
the experimental search at the PANDA experiment \cite{panda}, where the
glueball can be directly formed in proton-antiproton fusion process.

\subsection{Other glueballs}

A similar program can be carried out for all other glueballs in Table 7. For
instance, a tensor glueball with a mass of about $2.2$ GeV is expected to
exist from lattice calculations \cite{mainlattice}. A preliminary study of the
tensor glueball was presented long ago \cite{tensormio}. It would be
interesting to study this hypothetical state by using the eLSM presented
above. Another interesting channel is the vector one. Namely, a vector
glueball can be directly formed at BES.

In the future PANDA experiment \cite{panda} glueballs can be directly formed
in all non-exotic channels. A clear theoretical understanding of the decay
properties of glueballs will be also helpful for their future experimental search.

\section{Four-quark objects}

\subsection{ Four-quark states in the low-energy domain}

The light scalar resonances $f_{0}(500)$, $f_{0}(980)$, $a_{0}(980)$, and
$k=K_{0}^{\ast}(800)$ are not part of the eLSM. In fact, the attempt to
include them as ordinary quark-antiquark states shows that this scenario does
not work \cite{nf2,dick}. Then, as various other works confirm, these mesons
are not quark-antiquark states.

An elegant possible explanation of the nature of these mesons was put forward
long ago in Ref. \cite{jaffeorig,exotica} and further investigated in Refs.
\cite{maianilow,tq}, in which they are described as bound states of a good
diquark and a good anti-diquark. These diquarks are particularly stable since
the corresponding channel is attractive. An example of a wave-function of a
good diquark is given by:
\begin{equation}
\left\vert \text{space: }L=0\right\rangle \left\vert \text{spin:
}S=0\right\rangle \left\vert \text{color: }RG-GR\right\rangle \left\vert
\text{flavor: }ud-ds\right\rangle \text{ }%
\end{equation}
with $J^{P}=0^{+}$. For $N_{f}=3$ there are indeed 3 good diquarks:
\begin{equation}
\sqrt{\frac{1}{2}}[d,s]\text{ , }\sqrt{\frac{1}{2}}[u,s]\text{ , }\sqrt
{\frac{1}{2}}[u,d]\text{ ;}%
\end{equation}
(for an arbitrary $N_{f}$ there are $N_{f}(N_{f}-1)/2$ of such objects).

In this scenario one has:%
\begin{align}
f_{0}(500)  &  \equiv\frac{1}{2}[u,d][\bar{u},\bar{d}]\text{ ,}\\
f_{0}(980)  &  \equiv\frac{1}{2\sqrt{2}}([u,s][\overline{u},\overline
{s}]+[d,s][\overline{d},\overline{s}])\text{ ,}\\
a_{0}^{0}(980)  &  \equiv\frac{1}{2\sqrt{2}}([u,s][\overline{u},\overline
{s}]-[d,s][\overline{d},\overline{s}])\text{ ,}\\
k^{+}  &  \equiv\frac{1}{2}[u,d][\bar{s},\bar{d}]\text{ .}%
\end{align}
By taking into account the effective quark masses $m_{u}^{\ast}\simeq
m_{d}^{\ast}\simeq300$ MeV $<m_{s}^{\ast}$ $\simeq500$ MeV one can understand
the mass ordering%

\begin{equation}
M_{f_{0}(500)}<M_{k}<M_{f_{0}(980)}=M_{a_{0}(980)}%
\end{equation}
This is not possible for a nonet of conventional quark-antiquark states.
Moreover, also the decays can be correctly described \cite{tq}.

However, while such a tetraquark model is appealing, one should go beyond a
simple tree-level study and take into account quantum fluctuations. In fact,
all light scalar mesons are strongly affected by meson-meson loops. The
resonances $f_{0}(500)$ and $k$ are extremely broad ($\gtrsim400$ MeV), thus
the effect of quantum fluctuations is large \cite{lupo,zheng}. At the same,
the resonances $f_{0}(980)$ and $a_{0}(980),$ although not broad, are strongly
affected by the $\bar{K}K$ threshold, i.e. by kaon-kaon loops.

Indeed, various calculations could explain the light scalar mesons as
dynamically generated states \cite{lowscalars}. A very interesting possibility
is that the light scalar mesons emerge as companion poles of the heavier
quark-antiquark fields that we encountered in the previous section
\cite{tornqvist,pennington}. For instance, in the recent work of Ref.
\cite{thomaslast} it was possible to describe the quark-antiquark resonance
$a_{0}(1450)$ together with the dynamically generated state $a_{0}(980).$ in a
unified framework and with only one seed state. In this framework,
$a_{0}(980)$ does not survive in the large-$N_{c}$ limit: it simply fades out.

Summarizing, there is now a consensus that the light scalar mesons are
predominantly some sort of four-quark states, either as tetraquarks
\cite{jaffeorig,exotica,maianilow,tq} or in the form of molecular states
\cite{lowscalars} (for what concerns the state $f_{0}(500),$ we refer to very
recent review \cite{sigmareview} and refs. therein; for what concerns (the
indeed small) mixing with the ordinary mesons, see \cite{tqchiral,fariborz}).
Moreover, there is also an agreement among different studies about the fact
that mesonic interactions are crucial to understand these states. At the light
of the present evidence, the light scalar mesons should be considered as
non-conventional mesons.

As a last point, we would like to discuss how to incorporate the resonance
$f_{0}(500)$ in a chiral framework. In the case $N_{f}=2$ only one
tetraquark/molecular state is present, which we denote with the field $\chi.$
This object is chirally invariant. The coupling of $\chi$ to $\sigma$,
$\vec{\pi}$ and $G$ is described by the potential \cite{tqchiral}:%

\begin{equation}
V=V_{dil}(G)+aG^{2}(\sigma^{2}+\vec{\pi}^{2})^{2}+\frac{\lambda}{4}(\sigma
^{2}+\vec{\pi}^{2})^{2}-\sigma h_{0}+\frac{1}{2}m_{\chi}^{2}\frac{G^{2}}%
{G_{0}^{2}}\chi^{2}-g\frac{G}{G_{0}}\chi\left(  \sigma^{2}+\vec{\pi}%
^{2}\right)  \;\text{.} \label{vpot}%
\end{equation}
As soon as $G_{0}\neq0$, one has also that (for $a<0$) the field $\sigma$
condenses to $\sigma_{0},$ $\sigma_{0}\propto G_{0}\neq0,$ see Sec. Then, out
of the two terms $\frac{1}{2}m_{\chi}^{2}\chi^{2}+g\frac{G}{G_{0}}\chi
\sigma_{0}^{2}$ , we obtain also a v.e.v. for $\chi$:
\[
\chi_{0}\propto g\frac{\sigma_{0}^{2}}{m_{\chi}^{2}}\propto g\frac{G_{0}^{2}%
}{m_{\chi}^{2}}\text{ .}%
\]
In this simple model, a four-quark condensate emerges. Indeed, even this
simple system contains various mixing terms ($G$-$\sigma$, $\chi$-$\sigma$,
and $\chi$-$G$), which require numerical solutions. The field $\chi$ plays an
important role at nonzero temperature \cite{achim} and at nonzero density and
is also responsible of an attraction among nucleons \cite{susagiu,achimlast}.
Quite remarkably, even if $f_{0}(500)$ seems to play an important role for
explaining some feature at nonzero temperature and density, it is not
important in thermal models of heavy-ion collision. In fact, the contribution
of the scalar-isoscalar state $f_{0}(500)$ is nearly perfectly cancelled by
isotensor-scalar repulsion, see details in\ Ref. \cite{cancellation}.

\subsection{X,Y,Z states and other non-quarkonium candidates}

The discovery in the last years of many enigmatic resonances -so called
$X,Y,Z$ states- made clear that there are now many candidates of resonances
beyond the standard quark-antiquark scenario. We refer to Refs.
\cite{braaten,brambilla} for a list of the presently discovered states
($X(3872)$ was the first one experimentally found by BELLE in 2003). The
interpretation of these states is subject to ongoing debates: tetraquarks and
molecular interpretations are the most prominent ones, but it is difficult to
distinguish among them \cite{braaten,maianix}. At the same time, distortions
due to quantum fluctuations of nearby threshold(s) surely take place, thus
making a clear understanding of these objects more difficult \cite{coitox}. In
this respect, it would be interesting to perform a detailed study of the poles
on the complex plane of the new discovered resonances, in order to understand
if some of the newly discovered states are dynamically generated in the
meaning of Refs. \cite{thomaslast,pennington}.

There is however one aspect that should be stressed, since it shows that
objects beyond the quark-antiquark states have already been experimentally
found: these are the charged the $Z$ states, as for instance the state
$Z_{c}(3900)^{\pm}$ \cite{besz}. The mass of the state implies that this meson
contains a charm and an anticharm. Yet, how to make it charged? The only
possibility is to have four-quark states such as $c\bar{c}u\bar{d}\equiv
c\bar{c}\pi^{+}$ for $Z_{c}(3900)^{+}$ and $c\bar{c}d\bar{u}\equiv c\bar{c}%
\pi^{-}$ for $Z_{c}(3900)^{-}$ (e.g. Ref. \cite{maianiz}).

Although the exact configuration is not known, the evidence for a
non-conventional mesonic substructure is compelling. In addition, very
recently the neutral member of the multiplet $Z_{c}(3900)^{0}$ was found by
BES \cite{beszc0}, which corresponds naturally to $c\bar{c}(u\bar{u}-d\bar
{d})\equiv c\bar{c}\pi^{0}.$

In conclusion, we mention that there are other mesonic states which are not
yet understood. An example is the strange-charmed scalar state $D_{S0}(2317)$,
which is too light to be a $c\bar{s}$ quarkonium. It can be a four-quark or
-even more probable- a dynamically generated state.

\section{Conclusions}

In these lectures we have discussed one of the main topic of modern QCD: the
existence and the properties of non-conventional mesons, i.e. mesons which are
not quark-antiquark objects. Glueballs, i.e. bound state of gluons, are a firm
prediction of models and lattice QCD but are still missing detection, even if
some candidates exist. On the contrary, evidence toward the existence of
four-quark objects both at low-energy and in the charmonium region is mounting
(thanks to both experiment and theory).

For pedagogical purposes, we have first reviewed the symmetries of QCD.
Dilatation symmetry and its breaking is a central phenomenon of QCD since it
delivers a fundamental energy scale on which all quantities will depend.
Another very relevant symmetry is chiral symmetry and its spontaneous
breaking, which is responsible for the fact that chiral symmetry is hidden
(there are not degenerated chiral partners). Mass differences arises and
Goldstone bosons (most notably the pions) emerge. Another central theoretical
tool of QCD is the large-$N_{c}$ limit, in which the number of color is
artificially increased to large values. In this limit, conventional
quark-antiquark mesons as well as glueballs become stable.

Later on, we constructed a model of QCD, called the extended Linear Sigma
Model (eLSM), which embodies the mentioned properties: the presence of a
dilaton/glueball field is a primary ingredient; chiral symmetry and its
breaking are realized by a Mexican-hat potential; scalar, pseudoscalar,
vector, and axial-vector mesons are the building blocks of its Lagrangian. The
agreement of the theoretical results with data is very good (see Table 6 and
Ref. \cite{dick}).

The eLSM can be used to study glueballs. The scalar glueball is already there:
a detailed study of it has shown that the resonance $f_{0}(1710)$ is
predominantly gluonic. The pseudoscalar glueball has been also coupled to
light mesons and its branching ratios are predictions that can be used in
future searches of this particle. The very same idea can be extended to all
other glueballs predicted by lattice QCD. The eLSM offers a solid basis for
the calculation of the decays of gluonic states in accordance with chiral and
dilatation symmetries. In this context, the future PANDA experiment will be
very useful because almost all glueball can be produced in proton-antiproton
fusion processes.

Then, we have concentrated on four-quark objects. They can emerge as
diquark-antidiquark states and/or as meson-meson bound states. In any case, we
expect that quantum mesonic loops play an important role in the understanding
of various tetraquark candidates. In the low-energy sector, the light scalar
mesons are very broad and the use of quantum field theoretical models going
beyond tree-level is compelling. In the charmonium sector, many of the newly
discovered $X,Y,Z$ states are close to energy thresholds, thus also in this
case loops are expected to be relevant. In any case, it must be stressed that
the experimental discovery of charged $Z$ states implies that non-quarkonium
objects exist.

In conclusion, there is room for new discoveries in the field of
non-conventional hadrons. The interchange of experimental activity and
theoretical models and numerical simulations will be important to proceed in
this fascinating field of high energy physics.

\bigskip

\bigskip

\textbf{Acknowledgments: }I thank D.\ Parganlija\textbf{, }S. Janowski, T.
Wolkanowski, A. Heinz, S. Gallas, W. Eshraim, A. Habersetzer, Gy.\ Wolf, P.
Kovacs, and D. H. Rischke for collaborations on the topics discussed in this work.

\end{document}